\documentclass[]{interact}

\usepackage[numbers]{natbib} 
\usepackage{xcolor}
\usepackage{amsmath} 
\usepackage{graphicx}
\usepackage{tabulary}
\usepackage{tabularx}
\usepackage{makecell}
\usepackage{booktabs}
\usepackage{multirow}
\usepackage{multicol}
\usepackage{hyperref}
\usepackage{algorithm}
\usepackage[noend]{algpseudocode}
\usepackage{algorithmicx}
\usepackage{float}
\usepackage{nomencl}
\usepackage{etoolbox}
\usepackage{rotating}
\usepackage{tikz}
\usepackage{subcaption}
\usepackage{pdflscape}
\usepackage{afterpage}
\usepackage{soul}
\DeclareUnicodeCharacter{2212}{-}

\DeclareMathOperator*{\argmin}{argmin}
\newcommand{\Agents}{\mathcal{A}}
\newcommand{\Spots}{\mathcal{S}}
\newcommand{\Obs}{\mathcal{O}}
\newcommand{\Aenter}{\Agents_{\mathrm{enter}}}
\newcommand{\Aexit}{\Agents_{\mathrm{exit}}}
\newcommand{\Aother}{\Agents_{\mathrm{other}}}
\newcommand{\Sempty}{\Spots_{\mathrm{empty}}}
\newcommand{\Occ}{\mathbb{O}}
\newcommand{\ofi}{^{[i]}}
\newcommand{\ofs}{^{[s]}}
\newcommand{\ofo}{^{[o]}}
\newcommand{\R}{\mathbb{R}}

\newcommand{\Bspot}{\mathbb{B}_{\mathrm{spot}}}
\newcommand{\Bobs}{\mathbb{B}_{\mathrm{obs}}}
\newcommand{\mr}{\mathrm}

\def\tsc#1{\csdef{#1}{\textsc{\lowercase{#1}}\xspace}}
\tsc{WGM}
\tsc{QE}
\tsc{EP}
\tsc{PMS}
\tsc{BEC}
\tsc{DE}

\begin{document}


\title{Parking of Connected Automated Vehicles: Vehicle Control, Parking Assignment, and Multi-agent Simulation}

\author{
\name{Xu Shen\textsuperscript{1}, Yongkeun Choi\textsuperscript{1}, Alex Wong\textsuperscript{2}, Francesco Borrelli\textsuperscript{1}, Scott Moura\textsuperscript{3}, Soomin Woo\textsuperscript{3,4}\thanks{Corresponding Author. Email: soominwoo@konkuk.ac.kr}}
\affil{\textsuperscript{1}Mechanical Engineering, University of California, Berkeley, USA; \textsuperscript{2}Computer Science, University of California, Berkeley, USA; \textsuperscript{3}Civil and Environmental Engineering, University of California, Berkeley, USA;  \textsuperscript{4}Civil and Environmental Engineering, Konkuk University, Seoul, South Korea;}
}
 

\maketitle

\begin{abstract}
This paper introduces a comprehensive approach to optimize parking efficiency for connected and Automated vehicle (CAVs) fleets. We present a multi-vehicle parking simulator, equipped with hierarchical path planning and collision avoidance capabilities for individual CAVs. The simulator is designed to capture the key decision-making processes in parking, from low-level vehicle control to high-level parking assignment, and it enables the effective assessment of parking strategies for large fleets of ground vehicles. We formulate and compare different strategic parking spot assignments to minimize a collective cost. While the proposed framework is designed to optimize various objective functions, we choose the total parking time for the experiment, as it directly reflects the congestion level and the cost associated with the parking efficiency. We validate the effectiveness of the proposed strategies through an empirical evaluation against a dataset of real-world parking lot dynamics, realizing a substantial reduction in parking time by up to 43.8\%. This improvement is attributed to the synergistic benefits of driving automation, the utilization of shared infrastructure state data, the exclusion of pedestrian traffic, and the real-time computation of optimal parking spot allocation.
\end{abstract}


\section{Introduction}

\subsection{Background} 
Among the many aspects of modern life that demand our time and attention, parking is an often-overlooked but crucial aspect that directly affects the efficient flow of traffic and the quality of our lives.
A study by INRIX Research~\cite{graham_cookson_impact_2017} found that Americans waste an average of 17 hours per year searching for parking, at a total cost of \$73 billion in lost time, fuel, and emissions. Beyond the economic impact, parking inefficiencies contribute to stress, traffic congestion, and environmental damage. Time spent circling for spaces and delays in entering or exiting parking facilities increase fuel consumption and greenhouse gas emissions, degrading air quality and urban livability.

Addressing this pressing issue has motivated the development of various technologies aimed at reshaping parking infrastructure~\cite{banzhaf_high_2017, banzhaf_future_2017, serpen_design_2019, choi_optimal_2023}, optimizing resource allocation~\cite{geng_new_2013}, and refining pricing mechanisms~\cite{kotb_iparker-new_2016}. Most existing research approaches this problem from a macro perspective~\cite{arnott_modeling_1999}, modeling entire vehicle fleets as service queues~\cite{tavafoghi_queuing_2019} to improve parking lot utilization and alleviate urban traffic congestion~\cite{micus_customer_2022}. 
Although these methods have proven effective in distributing parking demand, there is a need for microscopic modeling of the parking vehicles because parking depends on drivers circulating in search of available spaces, which leads to considerable time and fuel inefficiencies. Moreover, such abstractions ignore individual vehicle kinematics and geometric constraints, limiting their relevance to fine-grained parking operations where spatial precision and motion feasibility are essential. Tight spaces, constrained layouts, and the need for precise maneuvering to avoid pedestrians and obstacles all introduce delays in the parking process. 

With advances in autonomous vehicle technologies, self-driving cars are now capable of navigating urban environments and operating as robotaxis in daily life. They also demonstrate strong potential for performing precise parking maneuvers and substantially improving parking efficiency. Consequently, automated parking has attracted considerable attention from both academia and industry.

From a research perspective, studies on automated parking can generally be categorized into two directions. The first focuses on enhancing the intelligence of a single vehicle, enabling it to navigate within parking lots and complete parking maneuvers autonomously. 
Key technologies include perception algorithms for detecting vacant parking spaces~\cite{suhr_sensor_2014, suhr_automatic_2010} and trajectory planning algorithms that allow vehicles to maneuver efficiently into designated spots~\cite{zhang_autonomous_2019, kim_neural_2021, li_time-optimal_2016, li_unified_2015, kim_auto_2014}. 
The second line of work addresses local conflicts among multiple vehicles during parking, such as collaborative path planning~\cite{kessler, shen2023multi, shen2023reinforcement}, driver behavior prediction~\cite{li_hierarchical_2019, shen}, and decision-making for finding empty spaces under limited information and competitive conditions~\cite{li_game_2020}. 
These efforts primarily emphasize the ego-vehicle's intelligence and local coordination, while neglecting fleet-level performance metrics such as system throughput and overall spatial efficiency.

In the industrial domain, leading companies such as Waymo and Tesla have made remarkable progress in single-vehicle autonomy, particularly in perception, control, and self-parking capabilities.
However, challenges remain in coordinating the collective fleet behavior in dense or shared parking environments. 
An incident in San Francisco, where a large group of Waymo robotaxis repeatedly honked and maneuvered aimlessly in a parking lot at night, vividly illustrates this limitation~\cite{tung_waymo_2024}. 
Although individual vehicles demonstrated precise control, the lack of fleet-level coordination led to severe congestion and operational inefficiency. 
This highlights an important research and industrial gap, where future development must extend beyond single-vehicle performance toward fleet-level optimization, integrating vehicle coordination~\cite{siegel_v2x, guanetti_control_2018}, dynamic space allocation, and infrastructure communication to achieve scalable, efficient, and socially acceptable automated parking systems.
 
Our prior work~\cite{shen2020autonomous} explored fleet parking heuristics within a small parking lot setting, introducing a grid-based collision avoidance approach. However, it was constrained to a small parking lot with only incoming vehicles, leaving plenty of room for designing a system applicable to environments with more complex structures and diverse traffic behaviors.  

\subsection{Contribution} 
In this paper, we design a comprehensive system for automated fleet parking and control, offering the following key contributions:
 
\begin{enumerate}
    \item {Hierarchical multi-vehicle parking simulator}: We propose a novel approach to studying the problem of automated fleet parking by developing a comprehensive simulation framework that integrates algorithms at various levels, from the low-level control of individual vehicle path planning and tracking, the mid-level algorithms for efficient collision avoidance, to the high-level fleet-wide parking spot assignment. Notably, we incorporate real-world human parking maneuver data to assess the impact of a full CAV fleet in densely populated scenarios.
    \item {Design and analysis of spot assignment strategies}: We devise multiple strategies for parking spot assignment, including a data-driven learning method. With the proposed simulator and the actual driving data recorded from a parking lot, we analyze various strategies across diverse parking and traffic scenarios against the human behavior baseline. 
    \item {Insights for real-world application}: Compared to a real-world parking lot operation, we show that 1) vehicle automation and communication with parking infrastructure can significantly reduce parking time, and 2) with our proposed data-driven strategy for spot assignment, we can reduce the parking time for a sizable fleet by up to 43.8\% compared to the baseline of human driving. This empirical validation emphasizes the real-world applicability and benefits of our approach. 
\end{enumerate}

\subsection{Research Overview}

To analyze the problem of efficient fleet parking through automation and communication, this work follows three main stages of development: single-vehicle control, multi-vehicle simulation, and vehicle assignment strategy. Figure~\ref{fig:structure} provides an overview of the research framework and illustrates how these components correspond to different sections of the manuscript. We adopt a bottom-up approach to presenting the framework:

\begin{figure*}
\centering
\includegraphics[width=\textwidth]{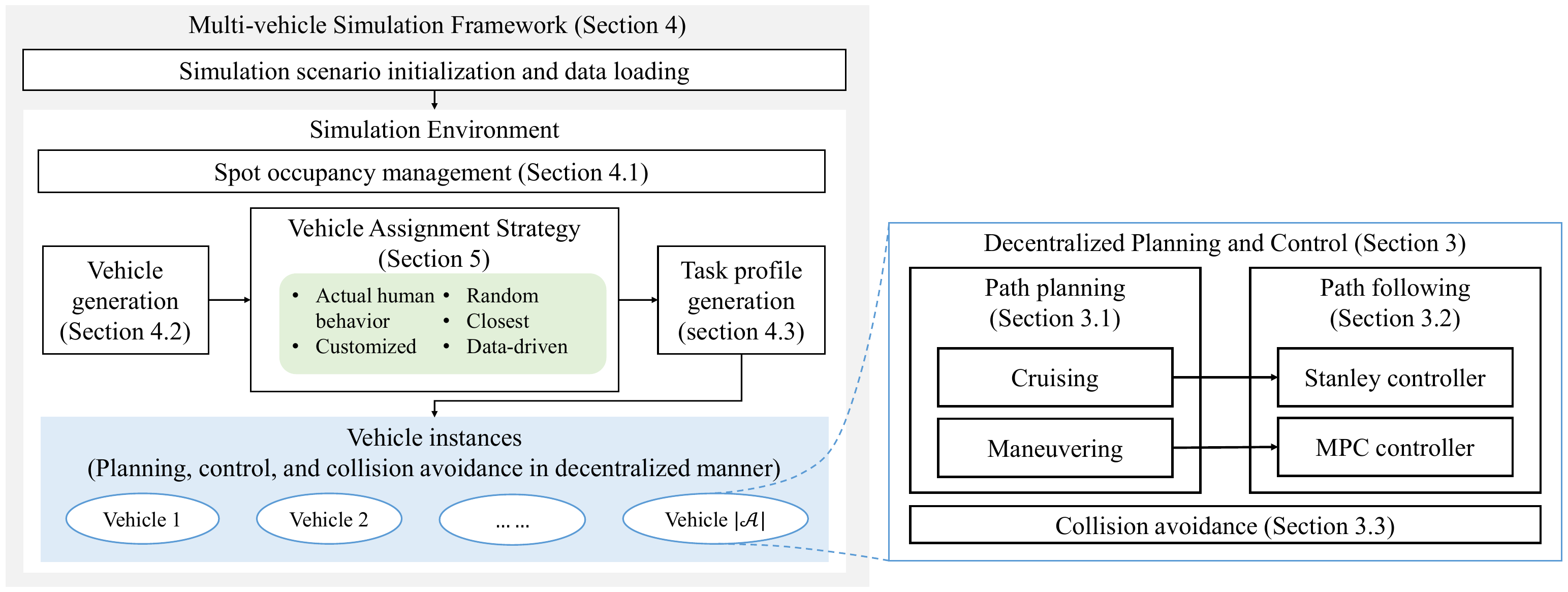}
\caption{Overview of the Simulator Framework and Control Strategies: The numbers indicate the sections in this manuscript that elaborate the design.}
\label{fig:structure}
\end{figure*}

In Section~\ref{sec:single-vehicle-planning-control}, we first introduce the control architecture of an individual vehicle. The vehicle begins by planning its path (Section~\ref{sec:pathplanning}), determining the route it will follow through the aisles of the parking lot (referred to as ``cruising'') and the trajectory for executing the movements to enter a parking spot (referred to as ``maneuvering''). It then follows the planned path (Section~\ref{sec:pathfollowing}) using distinct controllers for cruising and maneuvering modes. During path following, V2V communication is employed to ensure collision avoidance with nearby vehicles (Section~\ref{sec:collision-avoidance}). The process of leaving the parking lot follows a similar sequence in reverse.

The multi-vehicle simulator is presented in Section~\ref{sec:multi-vehicle-simulator}, which defines the environment where vehicles operate and interact. Vehicles are spawned at specified timestamps and locations (Section~\ref{sec:vehicle_spawning}). A spot occupancy manager ensures that multiple vehicles do not attempt to occupy the same parking spot simultaneously (Section~\ref{sec:spotoccupancymanagement}). According to the origin and destination, a sequence of tasks (referred to as ``task profile'') is assigned to define high-level behaviors to be executed in order (Section~\ref{sec:taskprofile}).

Finally, in Section~\ref{chap:vehicleassignment}, we propose and evaluate several parking spot assignment strategies, comparing their performance in terms of total driving time and overall efficiency. Although not explicitly illustrated in Figure~\ref{fig:structure}, we validate the proposed simulation framework and conduct extensive experiments to analyze the performance of these strategies in Section~\ref{chap:experiments}.

\section{Methodology}
In this section, we describe our formulation of the fleet parking problem, the assumptions on communication and computing technologies, and the real-world data from an actual parking lot with human-driven vehicles.

\subsection{Problem Definition}


A parking lot contains a set of parking spots $\Spots$, a set of static vehicles (referred to as obstacles) $\Obs$, a set of moving vehicles $\Agents$, and an entrance $e$ with coordinates $(x_e,y_e) \in \R^2$.  A parking spot $s \in \Spots$ is defined by its center $(x\ofs,y\ofs) \in \R^2$ and its width and length $(w\ofs,l\ofs) \in \R^2$. Its bounding box is defined by $\Bspot\ofs$. In this paper, we only work with parking spots that are perpendicular to the driving lanes, so the heading angle of each parking spot is ignored.
The set of unoccupied parking spots as $\Sempty = \{s\mid s\in \Spots,\Occ\ofs=0\}$.
The obstacles in $\Obs$ are vehicles that do not move for the entirety of the experiment time, so they are treated as static. An obstacle $o \in \Obs$ is defined by its center $(x\ofo,y\ofo) \in \R^2$, its width and length $(w\ofo,l\ofo) \in \R^2$, and its heading angle $\psi\ofo \in \R$. Its bounding box is defined by $\Bobs\ofo$.

Vehicles in $\Agents$ will move during the experiment time. The state $z\ofi(t)$ of a vehicle $i \in \Agents$ at time $t$ is defined by its center position $(x\ofi(t),y\ofi(t)) \in \R^2$, its width and length $(w\ofi,l\ofi) \in \R^2$, its heading angle $\psi\ofi(t) \in \R$, and its speed $v\ofi(t)$. Its bounding box is defined by $\mathbb{B}\ofi$. Each vehicle also has a time $t_{\mathrm{start}}\ofi$ when it enters the parking lot. For brevity, we group the fleet into disjoint sets: $\Aenter$ are vehicles that start at the entrance $e$ and end in a parking space, $\Aexit$ are vehicles that start in a parking space and exit at the entrance $e$, and $\Aother$ are vehicles with other types of behaviors such as switching spots or unloading. For each vehicle $i\in \Agents$ at time $t$, the control commands drive the vehicle along the path consisting of acceleration $a\ofi(t)$ and steering angle $\delta\ofi(t)$.

Denote by $f\ofi(t)$ the generic cost of vehicle $i\in \Agents$ at time $t$, the objective is to minimize this cost of all vehicles
\begin{equation}
    \sum_{i\in \Agents_{\mathrm{enter}}} \sum^{t\ofi_{\mathrm{end}}}_{t = t\ofi_{\mathrm{start}}} f\ofi(t),  
\end{equation}
where $t\ofi_{\mathrm{end}}$ is the time vehicle $i$ finishes its maneuver and stops inside a parking spot.

The generic cost, $f\ofi(t)$, can reflect any metrics of parking efficiency, including but not limited to time, passenger comfort, energy consumption, and emission. In the remainder of this paper, we will focus on time, i.e., $f\ofi(t) = 1$ unit of discrete time steps, since time is a straightforward representation of parking efficiency and is correlated to energy consumption~\cite{graham_cookson_impact_2017}.




Since vehicles will have complex behavior and interactions in the parking lot, an exact solution to this optimization problem is intractable. 
Instead, we address this problem by simulating the movement of a large fleet in a real-world parking lot and analyzing the effect of motion control design and different spot assignment strategies on the final cost. 

\subsection{Assumptions}
We make assumptions about communication and information availability to develop our parking controller as follows. First, we assume Vehicle-to-Infrastructure (V2I) communication, where each vehicle has information about the map of the parking lot, including the driving lanes and parking spot locations. Second, we assume Vehicle-to-Vehicle (V2V) communication, where each vehicle has information about other vehicles in the lot (both static and moving), as well as information related to their autonomous navigation (acceleration, braking, reference trajectory). 

We also assume decentralized computing for the motion control commands of the vehicle because the computational burden would be substantial, especially as the number of vehicles grows. Moreover, it is more robust to changes in a dynamic parking lot than centralized computing, as an unexpected vehicle that enters the lot may drastically change the optimal path for all vehicles. Therefore, we assume that each vehicle computes its own trajectory independently in a decentralized manner, and the central infrastructure only acts as an information hub. 

\subsection{Data}

\begin{figure}
     \centering
     \begin{subfigure}[b]{.7\linewidth}
         \centering
         \includegraphics[width=\linewidth]{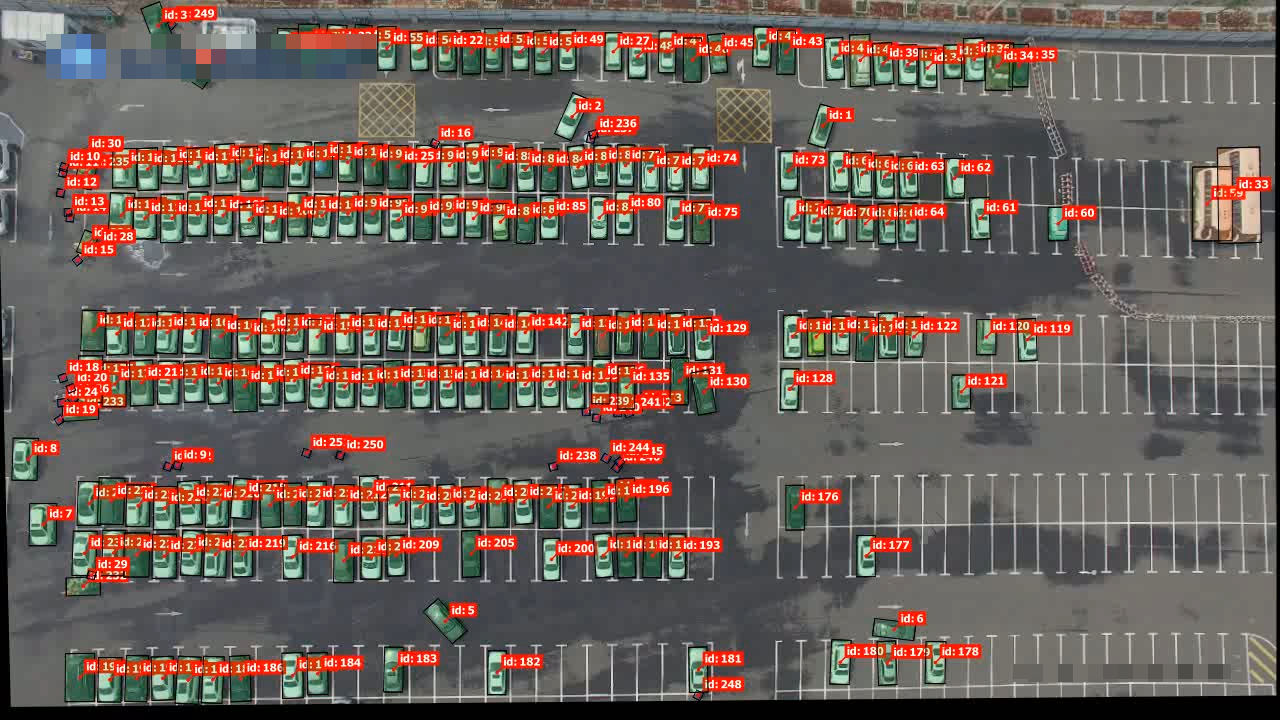}
         \caption{Sample frame from an annotated video}
         \label{fig:dlp}
     \end{subfigure}
     ~
     \begin{subfigure}[b]{.7\linewidth}
         \centering
         \includegraphics[width=\linewidth]{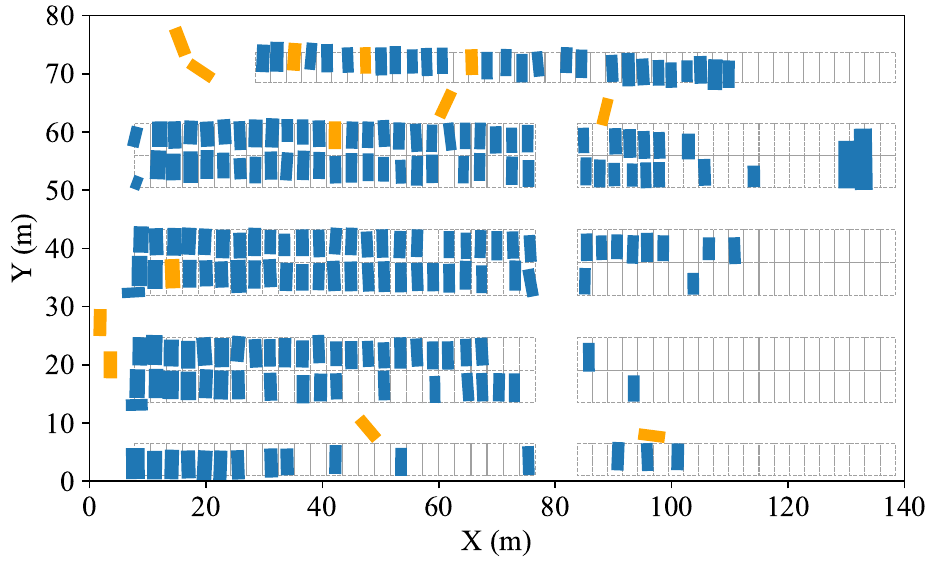}
         \caption{Visualization of the processed data}
         \label{fig:framesim}
     \end{subfigure} 
        \caption{Dragon Lake Parking (DLP) Dataset} 
\end{figure}
 
In developing our multi-agent simulation, we leverage the Dragon Lake Parking (DLP) dataset\footnote{\url{https://sites.google.com/berkeley.edu/dlp-dataset}}\cite{shen, shen_dragon_2023}, one of the largest publicly available datasets dedicated to parking scenarios. Collected over a continuous 3.5-hour period, the dataset contains detailed annotations of vehicles, pedestrians, and bicycles within a busy real-world parking lot. Figure~\ref{fig:dlp} illustrates a representative example of the annotated data.

In this work, the DLP dataset serves multiple purposes:
\begin{itemize}
    \item It provides the map of the parking lot, including geometric dimensions, parking spot locations, and a set of waypoints defining the driving lanes (``aisles'').
    \item The static vehicles in the dataset are used to define the obstacle set, $\Obs$.
    \item The starting times and locations of moving vehicles can be used to initialize fleet in the simulation.
    \item The observed spot selections in the dataset can serve as a baseline for evaluating various spot assignment algorithms.
    \item Performance metrics of human-driven vehicles in the data are used as a baseline for comparison.
\end{itemize}

Figure~\ref{fig:framesim} visualizes a sample frame from the dataset, where parking spots are drawn as dashed black lines, obstacles as blue rectangles, and vehicles as yellow rectangles.

\section{Decentralized Planning and Control Design} 
\label{sec:single-vehicle-planning-control}
This section describes the low-level algorithms that enable individual vehicles to cruise (i.e., navigate from one location in the parking lot to another) and to maneuver into and out of parking spots. The path planning process is detailed in Section~\ref{sec:pathplanning}, while feedback-based path following is introduced in Section~\ref{sec:pathfollowing}. A rule-based collision avoidance scheme is presented in Section~\ref{sec:collision-avoidance}.

\subsection{Path Planning} \label{sec:pathplanning} 

\subsubsection{Cruising} \label{sec:cruising}


\begin{figure}
     \centering
     \includegraphics[width=.7\linewidth]{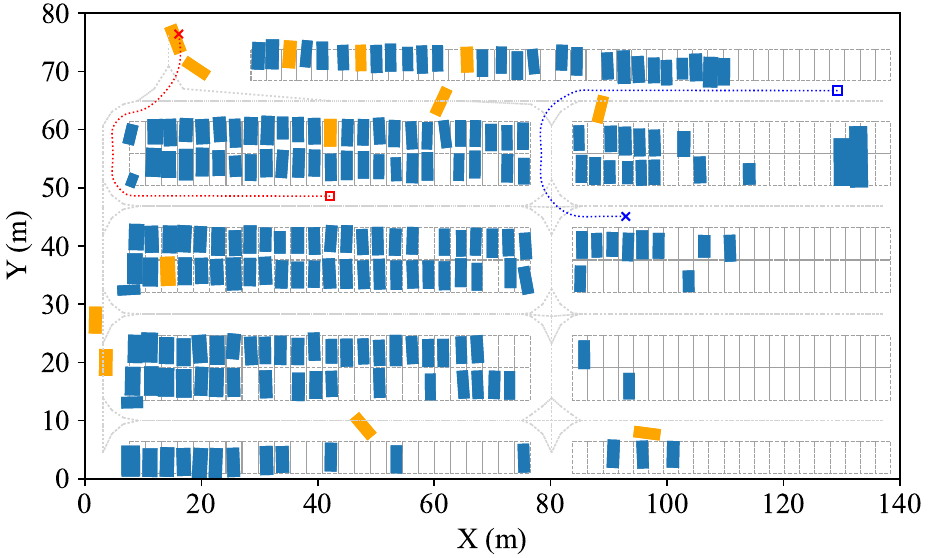}
    \caption{Path planning for cruising. In the driving lanes, the square markers indicate the starting location and the cross markers indicate the end location.} 
     \label{fig:astaroffset}
\end{figure} 

To enable vehicles to cruise along the aisles of the parking lot, we employ the A* search algorithm~\cite{hart} on the waypoint graph to compute the shortest feasible path between any given origin and destination. To further reduce the likelihood of collisions between vehicles traveling in opposite directions, the resulting paths are laterally shifted toward the right side of each aisle, following the designated driving direction. Figure~\ref{fig:astaroffset} illustrates several example cruising paths generated for different origin–destination pairs.




\subsubsection{Maneuvering} \label{sec:maneuvering}  

After a vehicle reaches its designated parking spot, it must depart from the aisle and maneuver into the spot, an operation that cannot be executed using the pre-defined waypoints along the driving lanes. Therefore, new paths must be generated to model the vehicle’s motion from the A* path terminus into the parking spot. Because parking environments typically offer limited clearance, these maneuvering paths must be both dynamically feasible and collision-free, ensuring that vehicles can follow them safely within tight spatial constraints.

\begin{figure}[h]
\centering
\includegraphics[width = .8\linewidth]{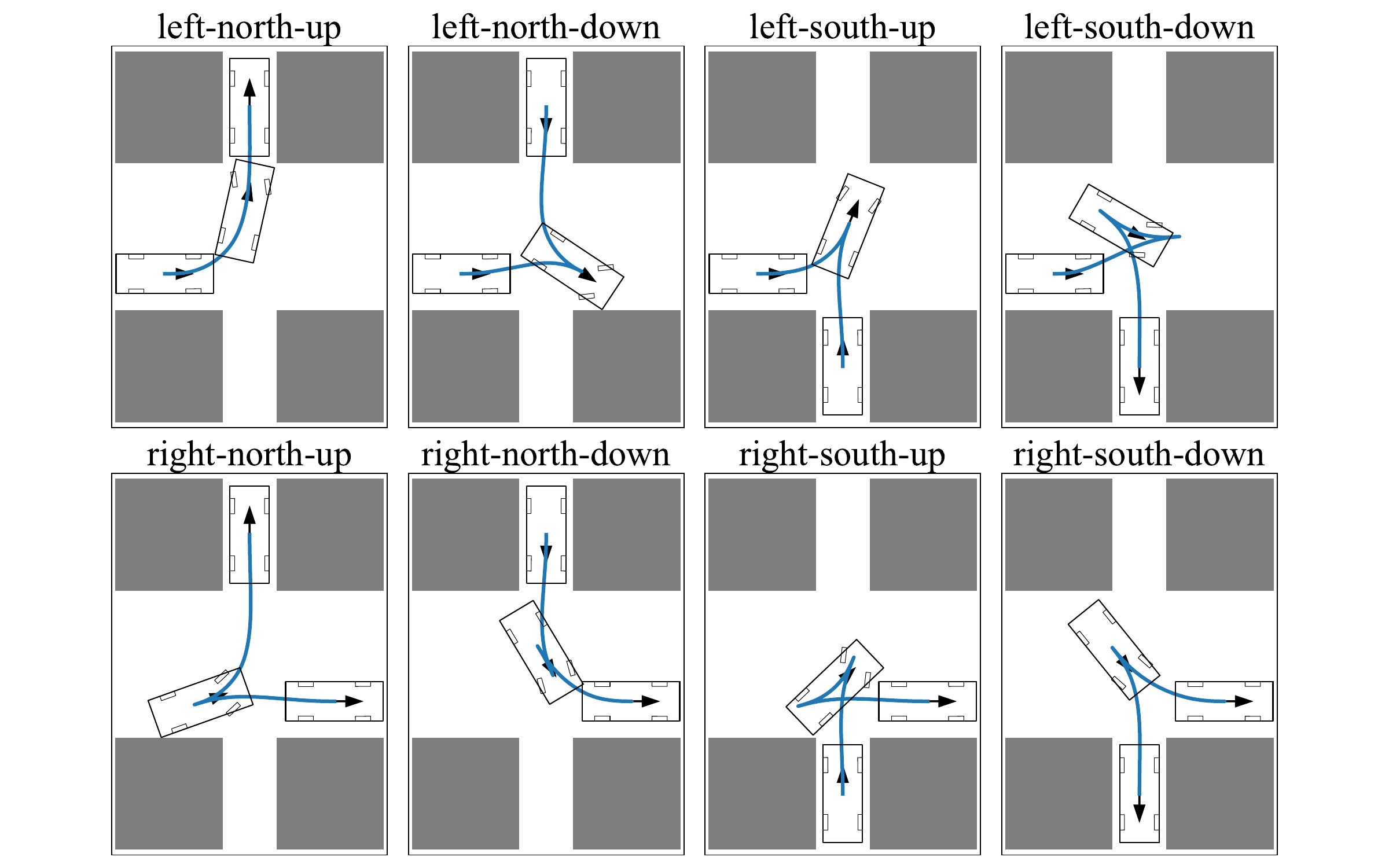}
\caption{Parking maneuver trajectories}
\label{fig:maneuver}
\end{figure}

Since all parking spots within a given lot share a similar orientation, the entry and exit maneuvers can be precomputed and reused across different spots. In other words, these paths can be generated once offline for each parking lot and applied as templates during simulation. For the DLP dataset, we construct eight distinct maneuver sets, referred to as ``parking maneuvers'', as illustrated in Figure~\ref{fig:maneuver}. Each maneuver is characterized by three criteria:
(1) the starting location (left or right), indicating the side of the spot from which the vehicle begins its maneuver;
(2) the target spot orientation (north or south), specifying the direction of the parking spot relative to the aisle; and
(3) the final heading (up or down), representing the desired vehicle orientation upon completing the maneuver.

To calculate the path for a given maneuver, we define the vehicle's state at the start and end of the maneuver to be $z_{0}$ and $z_{\mathrm{F}}$, respectively, and use the kinematic bicycle model $\dot{z} = f(z, u)$ as the vehicle dynamics:
\begin{equation}
    \dot{z} = f(z, u) := \left[
        \begin{matrix}
            \dot{x} \\ \dot{y} \\ \dot{\psi} \\ \dot{v}
        \end{matrix}
    \right] = \left[
        \begin{matrix}
            v\cos(\psi) \\ v\sin(\psi) \\ \frac{v}{l_{\mathrm{wb}}}\tan(\delta) \\ a
        \end{matrix}
    \right], \ u = \left[
        \begin{matrix}
            a \\ \delta
        \end{matrix}
    \right].
    \label{eq:kin_model_ct}
\end{equation}

\begin{figure} 
\centering
\includegraphics[width = .9\linewidth]{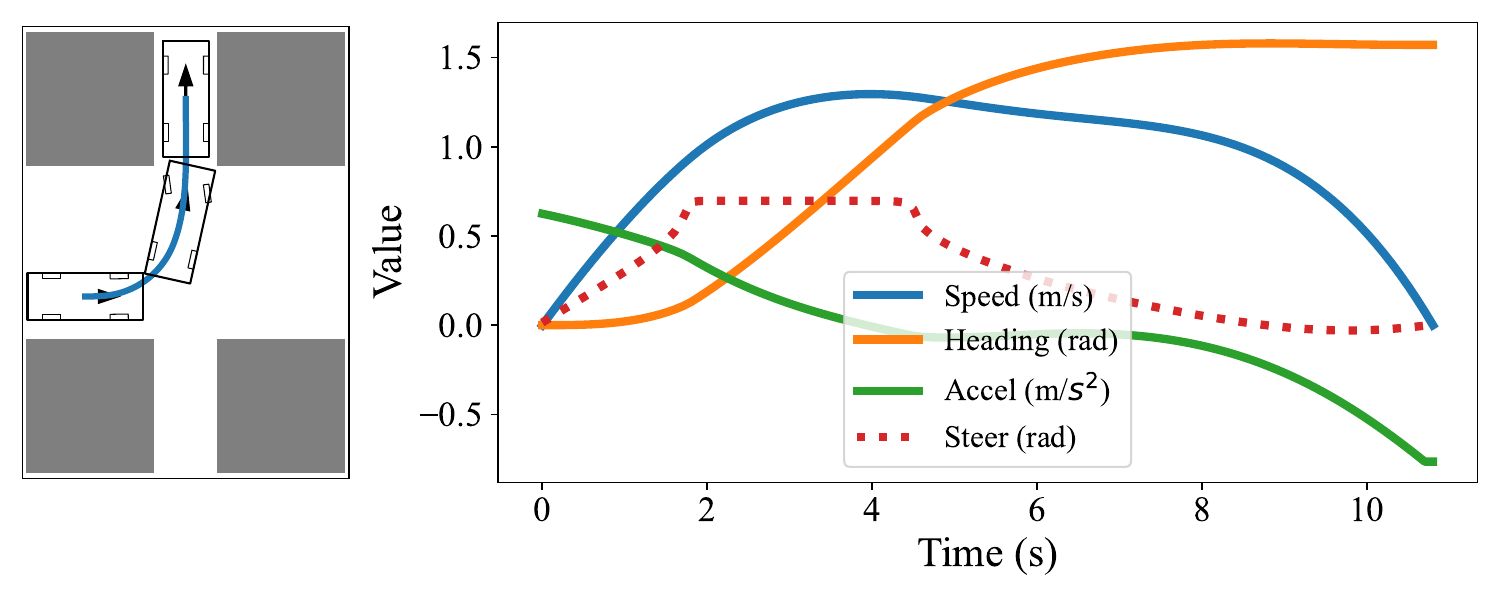}
\caption{A sample trajectory of ``left-north-up'' maneuver and corresponding state-input profiles.}
\label{fig:leftnorthup}
\end{figure}

Then, we find the optimal path and corresponding control commands by solving the following trajectory planning problem~\cite{obca}:
\begin{subequations}
\label{eq:obca}
\begin{align}
    \min_{\mathbf{z}, \mathbf{u}, T} \ & \ J = \int_{t=0}^{T} c\left(z(t), u(t)\right) dt\nonumber \\
    \text{s.t. } \ & \dot{z}(t) = f(z(t), u(t)), \\
    & z(t) \in \mathcal{Z}, u(t) \in \mathcal{U},\\
    & z(0) = z_0, z(T) = z_{\mathrm{F}}, \\
    &  \mathrm{dist}\left(\mathbb{B}(z(t)), \mathbb{B}^{[o]}_{\mathrm{obs}}\right) \geq 0, \forall o,
\end{align}
\end{subequations}
where the state $z$ and input $u$ are constrained under operation limits $\mathcal{Z}$ and $\mathcal{U}$. We denote by $\mathbb{B}(z(t))$ the vehicle body at time $t$ and ask it to maintain collision-free from all obstacles $o \in \mathcal{O}$. The stage cost $c(\cdot, \cdot)$ can encode the amount of actuation, energy consumption, and time consumption. The optimal solution $\{ \mathbf{z}^{*}, \mathbf{u}^{*} \}$ is the optimal path for the given maneuver. Figure \ref{fig:leftnorthup} shows a sample trajectory and the corresponding optimal inputs.

The eight maneuvers shown in Fig.~\ref{fig:maneuver} are sufficient to accommodate all possible parking configurations within our simulation. For vehicles driving along the other side of the aisle, corresponding maneuvers can be obtained simply by mirroring the existing trajectories. Similarly, to compute the trajectory for exiting a parking spot into an aisle, the same maneuver used for entry can be applied in reverse. Throughout this work, we refer to the process of entering a spot from an aisle as ``parking'' and the process of exiting a spot into an aisle as ``unparking''. Both operations are collectively categorized as maneuvering. 

\subsection{Path Following} \label{sec:pathfollowing}

After planning a vehicle's path by defining a set of waypoints, we use two algorithms to follow that path: a Stanley controller for the cruising path and an MPC controller for the maneuvering path.

\subsubsection{Stanley Controller} \label{sec:stanely-controller}

We use Stanley controller \cite{hoffmann}, a well-known real-time method to follow the path returned by our A* search in Sec.~\ref{sec:cruising}. Given a set of waypoints, it defines a control law to determine the vehicle's acceleration $a$ and steering angle $\delta$:

\begin{subequations} \label{eq:stanley}
\begin{align}
    a(t)=k_p(v(t)-v_{\text{ref}}), \\
    \delta(t)= \psi(t)+\arctan\frac{ke_{\mathrm{lat}}(t)}{v(t)},
\end{align}
\end{subequations}
where for each vehicle, $v$ is velocity, $v_{\text{ref}}$ is the reference velocity, $e_{\mathrm{lat}}$ is the lateral displacement error, and $\psi$ is the heading error of the vehicle with respect to the closest segment of the reference trajectory. 

\subsubsection{Model Predictive Control} \label{mpc-controller}

Although the Stanley controller provides efficient path-following performance, it does not account for collision avoidance. When driving through aisles, this limitation can be mitigated using a collision avoidance algorithm, as discussed in Section~\ref{sec:collision-avoidance}. During maneuvering, however, more precise control is required due to the proximity of parked vehicles.

To achieve accurate path following for parking and unparking maneuvers while maintaining reasonable computational cost, we employ Model Predictive Control (MPC). MPC minimizes deviation from a reference trajectory while satisfying dynamic and safety constraints by repeatedly solving an optimal control problem over a finite horizon. In our framework, the reference trajectory is taken from the offline parking maneuver described in Fig.~\ref{fig:maneuver}, and the constraints ensure collision-free motion with respect to other vehicles. The MPC formulation is defined as follows:

\begin{subequations}
\label{eq:MPC}
\begin{align}
    \min_{\mathbf{z}_{\cdot|t}, \mathbf{u}_{\cdot|t}} \ & \ J = \sum_{k=0}^{N-1} (x_{k|t}-x_{\mr{ref}}(k|t))^2+(y_{k|t}-y_{\mr{ref}}(k|t))^2 \nonumber \\
    \text{s.t. } \ & z_{k+1|t} = z_{k|t} + f(z_{k|t},u_{k|t})\Delta_t, \\
    & z_{k|t} \in \mathcal{Z}, u_{k|t} \in \mathcal{U},\\
    & z_{0|t} = z(t), \\
    &  \mathrm{dist}\left(\mathbb{B}(z_{k|t}), \mathbb{B}^{[o]}_{\mathrm{obs}}\right) \geq 0, \forall o,~\label{eq:CA-constr}\\
    & \forall k = 0, \dots, N-1, \nonumber
\end{align}
\end{subequations}
where $t$ is the current time relative to the beginning of the parking maneuver, $f$ is the kinematic bicycle model~\eqref{eq:kin_model_ct}, $\Delta_t$ is the time step size, and the state $z=[x,y,\psi,v]^T$ and input $u$ are constrained under operation limits $\mathcal{Z}$ and $\mathcal{U}$. As before, we denote by $\mathbb{B}(z_{k|t})$ the vehicle body at time $k$ and ask it to maintain collision-free from all obstacles $o \in \mathcal{O}$. The stage cost is solely dependent on the squared distance to the corresponding part of the parking maneuver, given by $z_{\mr{ref}}=[x_{\mr{ref}},y_{\mr{ref}},\psi_{\mr{ref}},v_{\mr{ref}}]^T$. The collision avoidance constraints are described as Equation (\ref{eq:CA-constr}), where $\mathrm{dist}(\cdot, \cdot)$ describes the signed distance between two convex polygons. A differentiable reformulation is included in the Appendix~\ref{app:obca}. After solving Eq.~\eqref{eq:MPC}, we apply the first input $u^{*}_{0|t}$ to the vehicle.

It is worth noting that the reference trajectories generated by Eq.~\eqref{eq:obca} are already optimized with respect to vehicle dynamics and collision avoidance.
Therefore, when computational resources are limited during simulation, vehicles can be simply interpolated along the reference trajectory using the precomputed control inputs, making use of the MPC controller only when there is a mismatch between planning and control dynamics, requiring real-time adjustment.


\subsection{Collision Avoidance} \label{sec:collision-avoidance}

Collision avoidance in parking lots fundamentally differs from standard road maneuvers. Unlike public roads, parking environments lack traffic signals and stop signs, and vehicle movement patterns are less structured, often involving unpredictable behaviors. Consequently, preventing collisions in such settings is inherently more complex. To address this, we design a decentralized, rule-based collision avoidance algorithm that promotes smooth and efficient traffic flow among autonomous vehicles.

The algorithm operates based on the following rules:
\begin{itemize}
    \item A vehicle yields when a potential collision with another vehicle is detected in the immediate future.
    \item At intersections, the vehicle that has progressed further into the intersection region has higher priority.
    \item Vehicles cruising in aisles have higher priority than those preparing to start a parking maneuver. Conversely, once a vehicle begins its parking or unparking maneuver, it gains priority over nearby vehicles. In other words, a vehicle must wait until its path is clear before initiating a maneuver, but once started, the maneuver proceeds without interruption until completion.
\end{itemize}

\begin{figure}
\centering
\includegraphics[width = .6\linewidth]{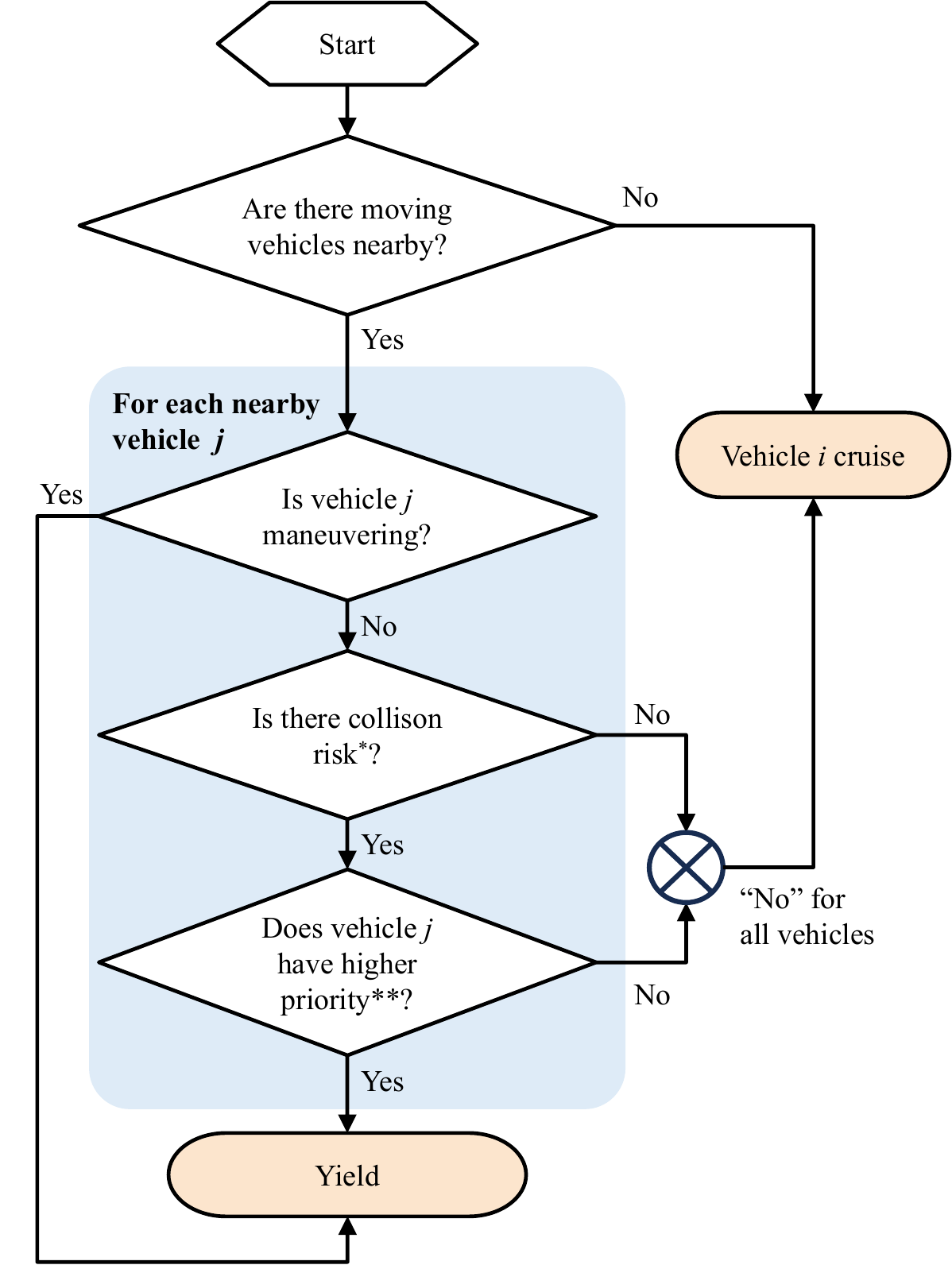}
\caption{Collision avoidance decision rules for a vehicle $i \in \Agents$ when cruising. $^*$The collision risk is evaluated with Algorithm \ref{alg:crashriskassessment}. $^{**}$The computation of priority is described in detail in Figure \ref{fig:shouldgobefore} and Algorithm \ref{alg:yieldassessment} in the Appendix~\ref{app:collavoid}.}
\label{fig:notbraking}
\end{figure}

\begin{figure} 
\centering
\includegraphics[width =.6\linewidth]{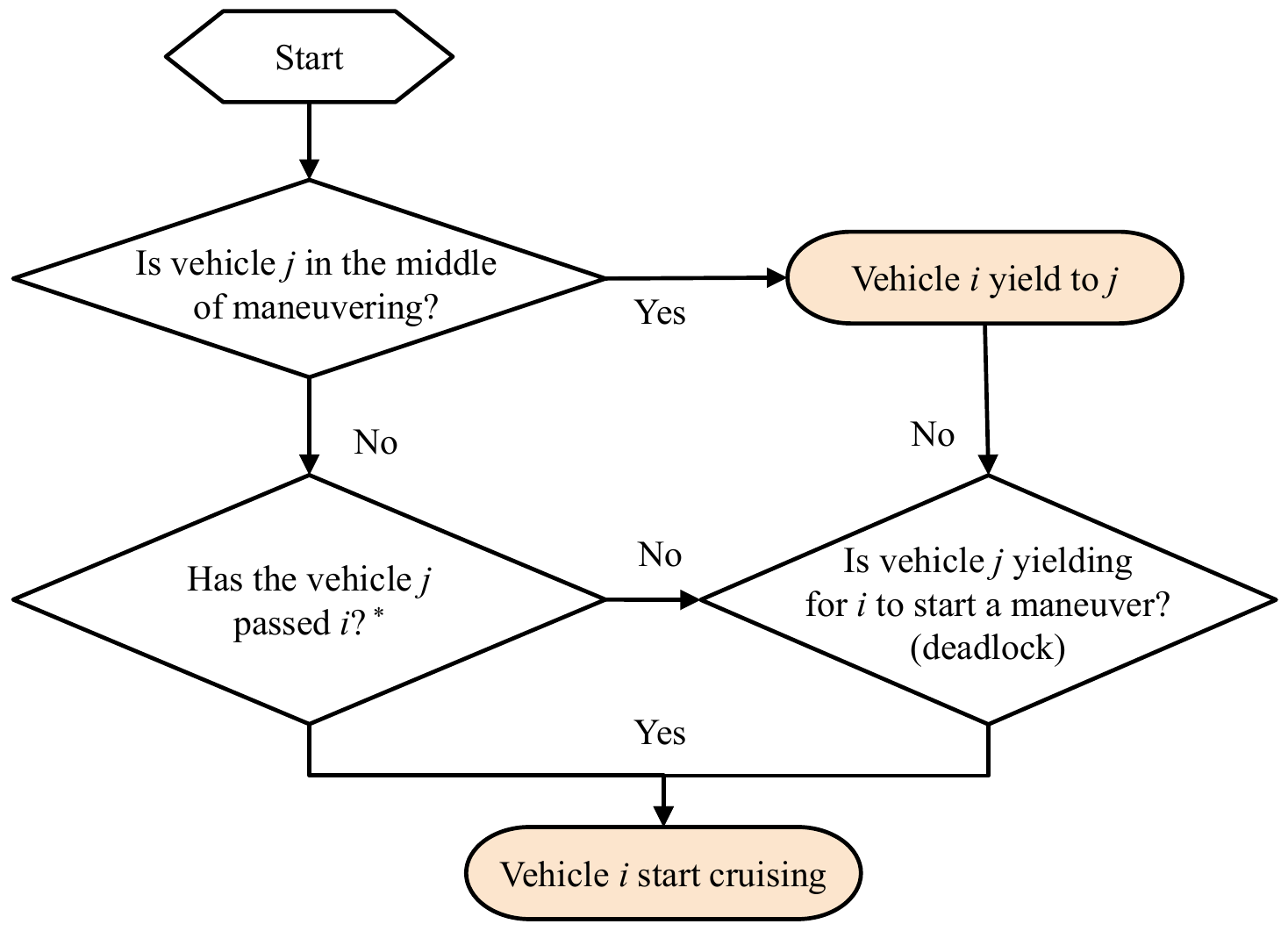}
\caption{Collision avoidance decision rules when a vehicle $i \in \Agents$ is yielding for $j \in \Agents / \{i\}$. $^*$ The criterion of ``$j$ passed $i$'' are described with more detail in Figure \ref{fig:haspassed} and Algorithm \ref{alg:passage_assessment} in the Appendix.}
\label{fig:braking}
\end{figure}

Based on these rules, each vehicle’s path-following behavior is governed by a finite state machine (FSM) with two states: cruising and yielding.
The decision rules that determine whether a vehicle remains in the cruising state or transitions to yielding are illustrated in Fig.~\ref{fig:notbraking}.
If a vehicle is already in the yielding state, it follows the rules in Fig.~\ref{fig:braking} to decide whether to remain stationary or resume cruising.

\section{Multi-Vehicle Simulator} \label{sec:multi-vehicle-simulator}

Building upon the decentralized planning and control framework described earlier, we develop a multi-vehicle simulation environment to evaluate fleet-level parking coordination.
The simulation process proceeds through the following stages:

\begin{enumerate}
    \item Initialization: The simulator initializes the scenario by loading map data, including parking spot locations and the positions of already parked vehicles.
    \item Vehicle spawning: Vehicles are dynamically generated in the environment (Section~\ref{sec:vehicle_spawning}) at entrances or vacant spots based on predefined or stochastic arrival patterns.
    \item Occupancy management: A central occupancy manager (Section~\ref{sec:spotoccupancymanagement}) acts as an info-hub to prevent over-assigning parking spots by maintaining real-time awareness of spot availability and vehicle intentions.
    \item Task assignment: Each vehicle is assigned a ``task profile'' (Section~\ref{sec:taskprofile}) -- a sequence of high-level behaviors such as cruising, parking, idling, or exiting, determined by the selected spot allocation strategy.
    \item Decentralized execution: After task assignment, vehicles plan and follow their own trajectories autonomously, avoiding collisions without further centralized control.
    \item Performance monitoring: The simulator continuously tracks the progress of all vehicles and records efficiency metrics for later analysis.
\end{enumerate}

\subsection{Vehicle Spawning} \label{sec:vehicle_spawning}

Vehicles can be spawned in the simulation environment using two approaches - first with predefined timestamps and locations and second with random arrival intervals at either the entrance or vacant parking spots. In the first approach, spawn times and locations can be derived from the DLP dataset or manually specified for controlled experiments.

In the second approach with random arrivals, we model the vehicles arrive at time intervals that follow an exponential distribution with mean $\lambda_{\mathrm{enter}}$.
In other words, once a vehicle $i \in \Aenter$ is spawned at time $t\ofi_{\mathrm{start}}$, the next entering vehicle $j \in \Aenter$ is spawned at $t^{[j]}_{\mathrm{start}}=t\ofi_{\mathrm{start}}+T_{\mathrm{interval}}$, where $T_{\mathrm{interval}}\sim \exp(\frac{1}{\lambda_{enter}})$. 
For those vehicles leaving the parking lot, they are spawned in random vacant parking spot, and start their unparking maneuvers with time intervals of the same distribution.

\subsection{Spot Occupancy Management} \label{sec:spotoccupancymanagement} 

To ensure vehicles do not compete over the same parking spot, the centralized infrastructure maintains an occupancy database $\Occ=\{0,1\}^{|\Spots|}$, where $\Occ\ofs$ is 1 if parking spot $s \in \Spots$ is occupied. 
We follow the rules in the following to update the database. Note that reusing the spots is  possible throughout a simulation.

\begin{enumerate}
    \item As the initial state of the simulation, for each $s \in \Spots$, $\Occ\ofs=1$ if there is an obstacle $o\in \Obs$ such that $(x\ofo,y\ofo)\in \Bspot\ofs$ or an existing vehicle $i\in \Agents$ such that $(x\ofi,y\ofi)\in \Bspot\ofs$;
    \item When a vehicle $i \in \Aenter$ is assigned to park in spot $s \in \Spots$, then $\Occ\ofs$ is set to 1;
    \item When a vehicle $i \in \Aenter$ finishes its maneuver out of the spot $s$,  $\Occ\ofs$ is set to 0.
\end{enumerate}

\subsection{Task Profile} \label{sec:taskprofile}
We define a ``task profile'' to represent the sequence of high-level driving behaviors assigned to each vehicle. The task profile of vehicle $i \in \Agents$, denoted as $\rho\ofi$, consists of an ordered list of tasks $\tau \in \rho\ofi$ selected from the set {\texttt{IDLE}, \texttt{CRUISE}, \texttt{PARK}, \texttt{UNPARK}}. For example, a vehicle that enters the parking lot, parks, waits, and then exits would have the task profile [\texttt{CRUISE}, \texttt{PARK}, \texttt{IDLE}, \texttt{UNPARK}, \texttt{CRUISE}] with appropriate parameters defined for each task, as summarized in Table~\ref{tab:tasks}.
During simulation, each vehicle executes its task profile sequentially, enabling seamless transitions between planning and control algorithms tailored to each stage of driving. 

\begin{table*}
\centering
\caption{Required parameters for each task profile}
\begin{tabular}{cccccc}
\hline
\makecell[c]{Task\\name} & \makecell[c]{Maximum\\driving speed,\\$v_{\mathrm{cruise}}$} & \makecell[c]{Spot to cruise\\to and park,\\$s_{\mathrm{target}}$} & \makecell[c]{Coordinates\\to cruise to,\\$x_{\mathrm{target}}, y_{\mathrm{target}}$} & \makecell[c]{Idle task\\duration,\\$\Delta T_{\mathrm{idle}}$} & \makecell[c]{Time to end\\idle task,\\$T_{\mathrm{next}}$} \\
\hline
\texttt{IDLE} & n/a & n/a & n/a & optional & optional \\
\texttt{CRUISE} & required & optional & optional & n/a & n/a \\
\texttt{PARK} & n/a & required & n/a & n/a & n/a \\
\texttt{UNPARK} & n/a & required & n/a & n/a & n/a \\
\hline
\end{tabular}
\label{tab:tasks}
\end{table*}

\section{Vehicle Assignment Strategy} \label{chap:vehicleassignment}
This section presents several strategies for assigning parking spots to a fleet of vehicles, with the goal of reducing local traffic congestion and improving overall parking efficiency.
By determining parking assignments at the entrance, each vehicle can be directed toward an optimal spot and follow an efficient route to its destination, minimizing unnecessary cruising within the lot. 

We evaluate multiple spot assignment algorithms, namely Human selection, Customized Solution, Random, Closest, and Data-driven algorithms, where the first two algorithms serve as baselines.
The first baseline is Human Selection, which represents the performance of human-driven parking behaviors observed in the real-world DLP dataset and is used to compare the algorithms to the status quo parking efficiency.
The second baseline is Customized Solution, which is a manually optimized configuration that approximates the upper bound of achievable performance, though it is generally impractical to compute automatically and generalize. This is used to compare the optimality of algorithms.

\subsection{Human Selection (Status Quo Baseline)} 

This strategy replicates the parking behavior of human drivers as recorded in the DLP dataset, representing the current, real-world status quo parking operation.
In this approach, the vehicle control and navigation are automated, but each entering vehicle is assigned the same parking spot it selected in the dataset.
This allows for a direct comparison between human parking choices and algorithmic spot assignment strategies under identical traffic and environmental conditions.

\subsection{Customized Solution (Algorithmic Baseline)} 

\begin{figure}[h]
\centering
\includegraphics[width=.7\linewidth]{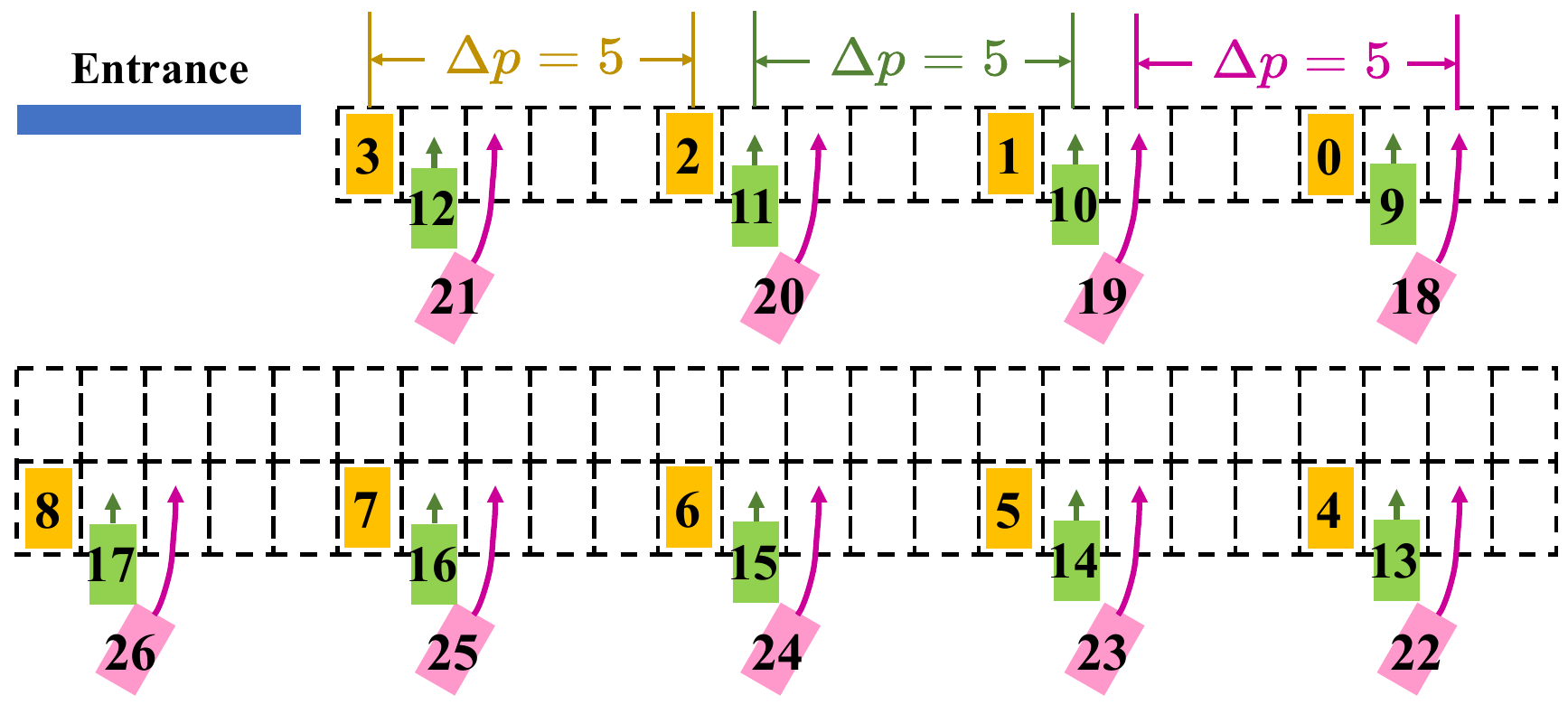}
\caption{An illustration of the Customized Solution assignment strategy. The parking spots are marked with dash lines. Vehicles are represented by solid rectangle boxes oriented towards their assigned parking spots. The numbers represent the order of arrival, i.e. number $0$ is the first vehicle, and number $26$ is the last vehicle. As indicated in orange, green, and purple colors, we will maintain a minimum distance of $\Delta p = 5$ for all consecutive assignments.}
\label{fig:handpicked}
\end{figure}

This strategy serves as an algorithmic baseline to benchmark the performance of other assignment methods, particularly in cases where human parking data are unavailable.
In this approach, the order of parking spot assignments is manually designed for a limited set of controlled scenarios.
As demonstrated by Shen et al.~\cite{shen2020autonomous}, maintaining adequate spacing between vehicles allows multiple parking maneuvers to occur in parallel without mutual obstruction, thereby improving overall fleet efficiency.
Accordingly, in this strategy, each pair of consecutive assignments is separated by a minimum distance $\Delta p$ while keeping the selected spots relatively close to the entrance.
For example, Figure~\ref{fig:handpicked} illustrates the strategy applied to an empty parking lot with three rows and $|\Aenter| = 27$, using $\Delta p = 5$.

However, this strategy is difficult to generalize into a fully automated algorithm, as it depends heavily on human judgment to adapt to varying parking lot geometries and occupancy conditions.
Moreover, applying this heuristic to partially occupied lots becomes particularly challenging, since the presence of pre-parked vehicles can easily disrupt the intended spatial pattern of assignments.

\subsection{Random}

In this algorithm, a spot is assigned at random from the set of all unoccupied spots $\Sempty$. We randomly sample an integer, $i$, from a discrete uniform distribution from one to the number of empty spots in the lot ($|\Sempty|$) and use the $i$-th spot in $\Sempty$.

\subsection{Closest}

This strategy assigns the arriving vehicle to the closest unoccupied spot from the entrance, i.e., 
\begin{equation}
    \hat{s} = \argmin_{s \in \Sempty} (x\ofs-x_e)^2 + (y\ofs-y_e)^2,
\end{equation}
where $(x_e, y_e)$ are the coordinates of the entrance.

\subsection{Data-driven} \label{sec:centralizednn}

This strategy aims to learn an estimator from data that predicts the total time consumption for parking and assigns each incoming vehicle to the spot with the lowest estimated time.
This algorithm attempts to model the parking time of vehicles that can be influenced by multiple factors, including the location of available parking spots and the level of real-time traffic congestion.

Given the potentially nonlinear relationship between input features and parking time, we employ a shallow neural network parameterized by $\theta$ to approximate this mapping: $g_{\theta}(\cdot): \mathbb{R}^{\mathcal{K}} \mapsto \mathbb{R}$. The model is trained using data collected from the DLP dataset, denoted as $\mathcal{D} = \left\{ (X\ofs, T\ofs_{\mathrm{gt} })\right\}$, where each record corresponds to a driver’s parking event. For each event, we extract the final parking spot $s$, the associated feature vector $X\ofs$, and the actual parking time $T\ofs_{\mathrm{gt}}$ as the ground-truth label. The feature vector is defined as
\begin{equation}
    X\ofs = \left[x\ofs, y\ofs, L\ofs, N\ofs_{\mathrm{path}}, N\ofs_{\mathrm{neighbor}}, \lambda_{\mathrm{enter}}, N_{\mathrm{queue}}\right]^{\top} \in \mathbb{R}^{\mathcal{K}},
\end{equation}
where $x\ofs, y\ofs$ denote the $(x, y)$ coordinates of parking spot $s$;
$L\ofs$ is the path length from the entrance to spot $s$ (as described in Section~\ref{sec:cruising});
$N\ofs_{\mathrm{path}}$ and $N\ofs_{\mathrm{neighbor}}$ represent the number of moving vehicles along the path and near the spot, respectively, capturing traffic-related effects;
$\lambda_{\mathrm{enter}}$ is the average vehicle arrival rate; and
$N_{\mathrm{queue}}$ is the number of vehicles waiting at the entrance.

The network is trained to minimize the Mean Squared Error (MSE) between predicted and actual parking times. After training, the optimal parameters $\theta^{}$ are used for online assignment, where each entering vehicle selects the parking spot with the minimum predicted parking time:
\begin{equation}
    \hat{s} = \argmin_{s \in \Sempty} g_{\theta^{*}}(s).
\end{equation}

\section{Evaluation of Parking Assignment Strategies} \label{chap:experiments}
This section evaluates the performance of the proposed simulator and the various parking assignment strategies introduced in Section~\ref{chap:vehicleassignment}.
We design two simulation scenarios:
\begin{enumerate}
    \item an initially empty parking lot with varying rates of randomly entering and exiting vehicles, and
    \item a partially occupied lot initialized with vehicle positions recorded in the DLP dataset.
\end{enumerate}

For the planner and controllers described in Eqs.~\eqref{eq:obca}, \eqref{eq:stanley}, and \eqref{eq:MPC}, we impose control input constraints of $|a| \leq 10~\mathrm{m/s}^2$ and $|\delta| \leq 40^{\circ}$.
According to the path-following state determined by the collision avoidance algorithm in Section~\ref{sec:collision-avoidance},
a vehicle in the cruising state is controlled using $v\ofi_{\mathrm{ref}} = v_{\mathrm{cruise}}$ and $k_p = 1$ in Eq.~\eqref{eq:stanley}, where $v_{\mathrm{cruise}}$ is adjusted based on the simulation scenario.
In the yielding state, the parameters are set to $v\ofi_{\mathrm{ref}} = 0$ and $k_p = 5$.

For the data-driven estimator described in Section~\ref{sec:centralizednn}, we use a multi-layer perceptron (MLP) with two hidden layers of sizes 84 and 10, employing ReLU activations.
The model is trained using the Adam optimizer with learning rate $= 0.01$.

All simulations are executed on an MSI Stealth 15M laptop equipped with a 12th Gen Intel Core i7-1280P CPU (2.00 GHz).
The simulator is implemented in Python 3.8 and is open-sourced at
\url{https://github.com/XuShenLZ/ParkSim}
.

\begin{table}
\centering
\caption{Parameter Set for the Empty Lot Experiments}
\begin{tabular}{cccc}
\toprule
\makecell{Parameter\\set} & \makecell{Number of\\entering vehicles} & \makecell{Number of\\exiting vehicles} & \makecell{Average Rate of \\ Entering\\or Exiting (s)} \\
\midrule
1 & 30 & 0 & 8\\
2 & 15 & 15 & 8\\
3 & 15 & 15 & 12\\
4 & 10 & 20 & 8\\
5 & 10 & 20 & 12\\ 
\bottomrule
\end{tabular}
\label{tab:parameterset}
\end{table}
 
\subsection{Simulation Scenario: Empty Lot}
\label{sec:empty_lot_result}
In this scenario, we begin with an empty parking lot and generate varying levels of entering and exiting traffic to evaluate the impact of different parking spot assignment strategies on fleet-level performance.

\begin{figure} 
\centering
\includegraphics[width = .7\linewidth]{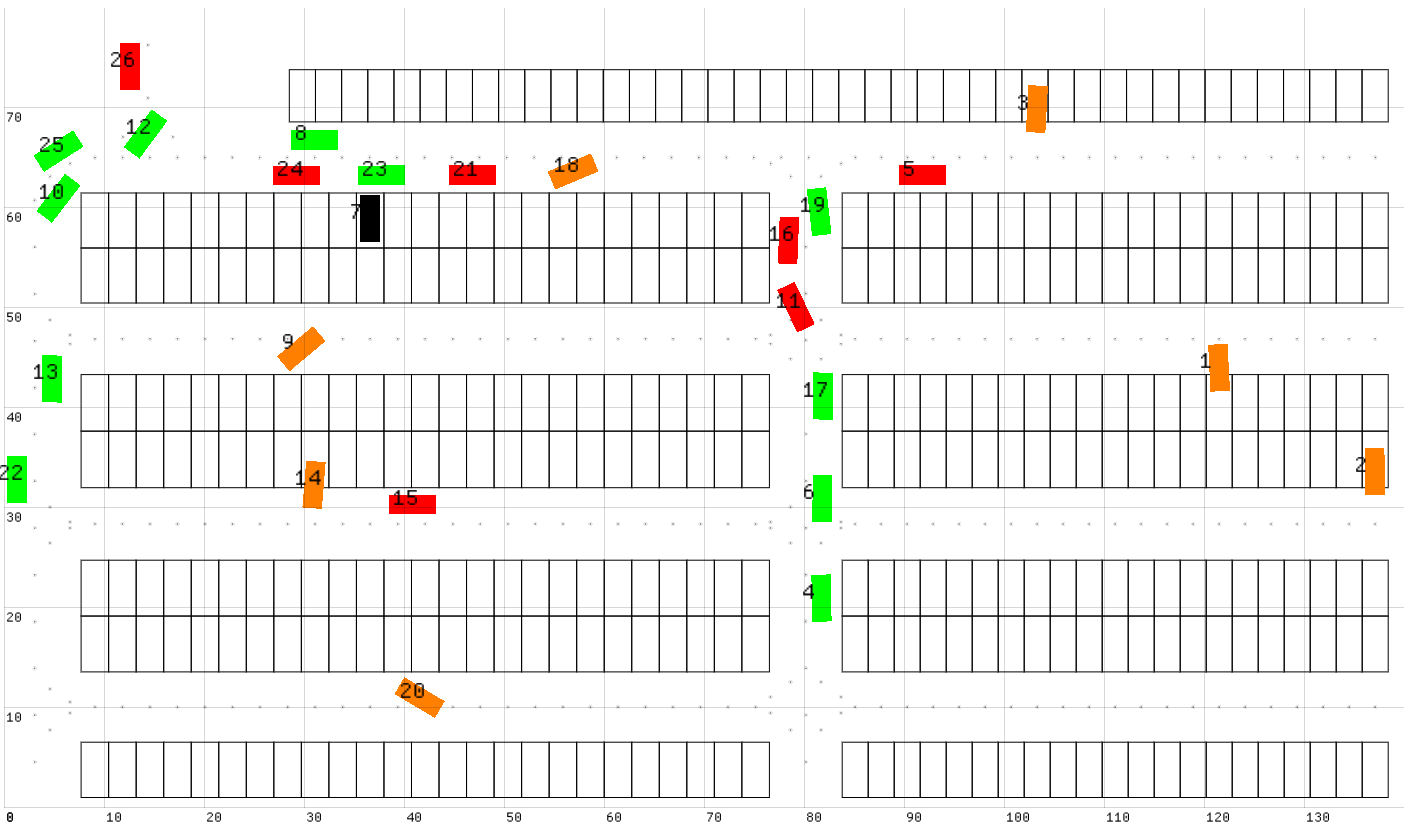}
\caption{Snapshot of the simulation in an empty parking lot. Vehicles are color-coded to indicate different driving states.}
\label{fig:sim_demo_empty_lot}
\end{figure}

Figure~\ref{fig:sim_demo_empty_lot} shows a representative simulation snapshot.
Vehicles are color-coded according to their current states:
green indicates cruising, red denotes yielding, orange represents active parking or unparking maneuvers, and black marks vehicles that have completed their tasks and remain stationary.

\subsubsection{Methodology}
We simulate traffic scenarios involving 30 vehicles under multiple parameter configurations summarized in Table~\ref{tab:parameterset}.
Parameter Set~1 models a case where all 30 vehicles enter the lot rapidly.
Sets~2 and~3 simulate balanced traffic, with an equal number of entering and exiting vehicles but different rates of entry and exit.
Sets~4 and~5 represent scenarios where fewer vehicles enter than exit, each with distinct entry and exit dynamics. 

Entering vehicles ($\Aenter$) are spawned at the lot entrance following exponentially distributed time intervals, while exiting vehicles ($\Aexit$) are generated randomly at the ends of driving aisles with exponentially distributed starting times.
All vehicles share a reference cruising speed of $v\ofi_{\mathrm{cruise}} = 5~\mathrm{m/s}$.

We evaluate four assignment strategies: {Customized Solution, Random, Closest, and Data-driven}. The Customized Solution serves as the baseline, as it approximates the global optimum when traffic congestion is minimal and multiple vehicles can maneuver simultaneously.
Each simulation scenario is repeated 10 times per assignment policy to account for stochastic variation.

Additionally, we conduct a sensitivity analysis on the vehicle arrival rate (3.75, 5, 7.5, and 15 vehicles/min) while keeping 30 entering and 10 exiting vehicles.
This tests each strategy’s robustness under various traffic inflows.
Each configuration is repeated 10 times, and results are reported as the average total driving time.

\begin{figure}
\centering
\includegraphics[width = .7\linewidth]{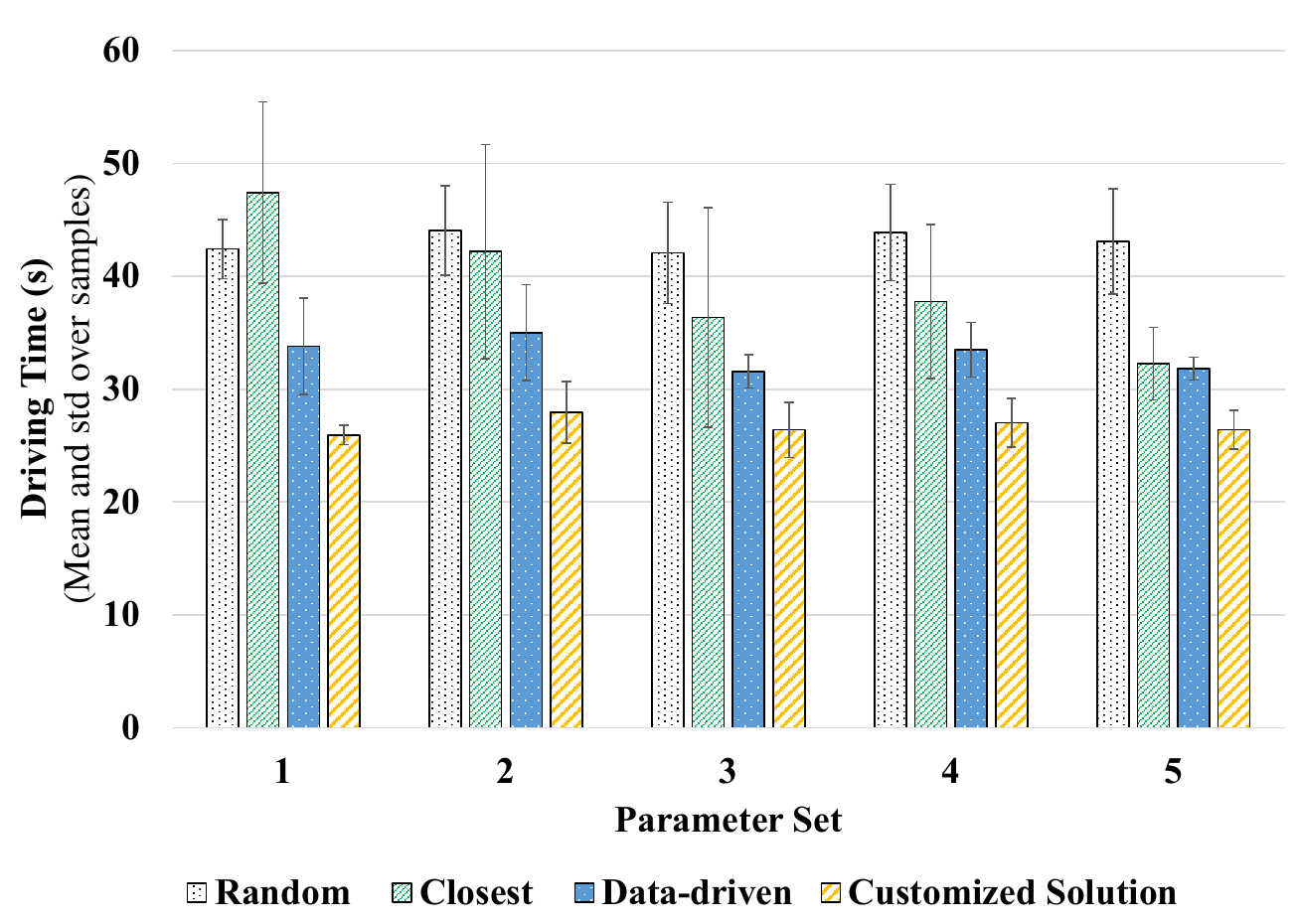}
\caption{Driving time under various vehicle assignment strategies in an empty parking lot with varying traffic parameter sets (see Table~\ref{tab:parameterset}).
The Data-driven strategy closely matches the Customized Solution baseline, outperforming Random and Closest strategies.}
\label{fig:maponly}
\end{figure}

\begin{figure} 
\centering
\includegraphics[width = .7\linewidth]{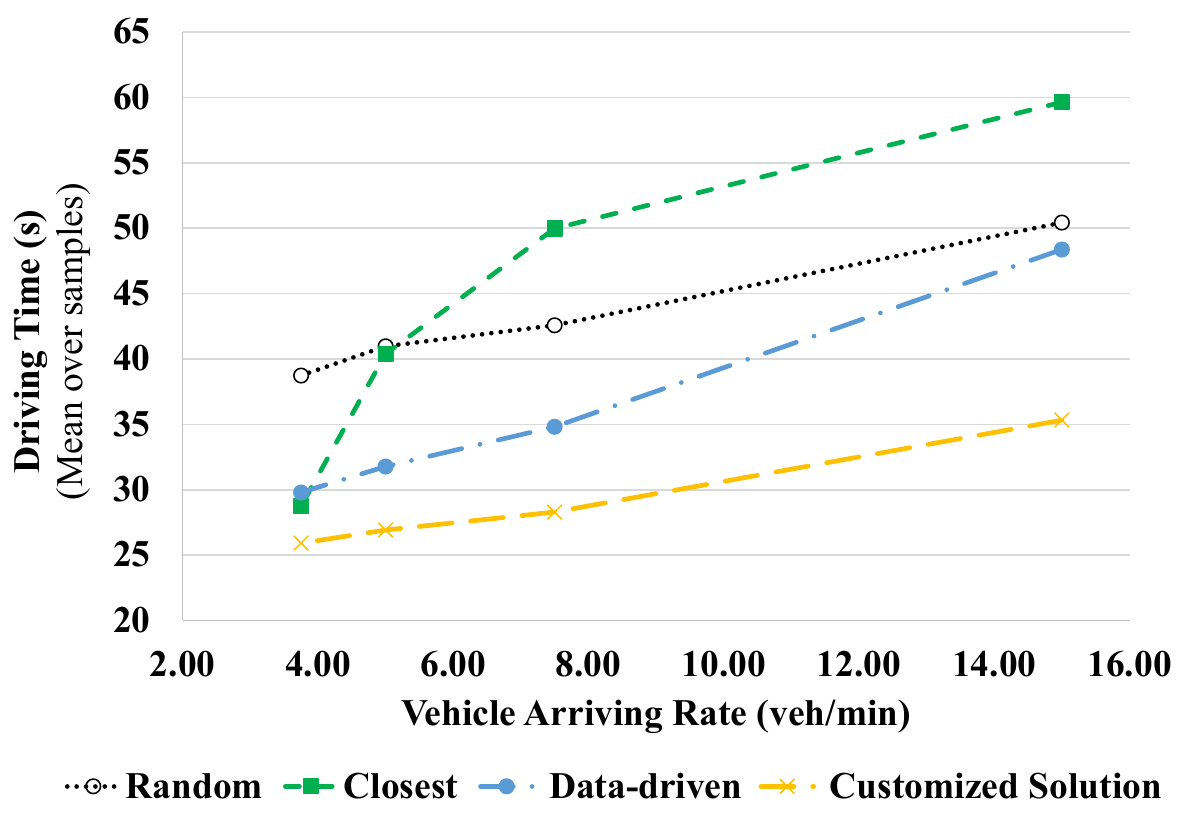}
\caption{Average driving time under different vehicle arrival rates (3.75–15 vehicles/min) with 30 entering and 10 exiting vehicles.
The Data-driven strategy achieves performance closest to the Customized Solution baseline.}
\label{fig:vehiclearrivingrate}
\end{figure}

\subsubsection{Results} 
Figure~\ref{fig:maponly} compares the performance of different assignment strategies across the parameter sets in Table~\ref{tab:parameterset}.
The {Random} strategy distributes vehicles roughly uniformly throughout the lot, leading to stable but suboptimal performance.
It often assigns vehicles to distant spots even when closer spaces are available.
The {Closest} strategy, in contrast, consistently selects nearby spots, which increases congestion and queuing near the entrance—especially at higher traffic densities.
It performs better only in low-density conditions with fewer entering vehicles.

The {Data-driven} method achieves performance closest to the Customized Solution baseline.
By learning spatial and traffic-dependent patterns, it assigns vehicles near the entrance while maintaining a balanced distribution across aisles, effectively avoiding localized congestion.
Certainly, the {Customized Solution} achieves the best performance across all configurations by manually spacing assignments to allow parallel parking maneuvers and minimal interference, though it is not scalable for real-world automation.

Figure~\ref{fig:vehiclearrivingrate} further examines the effect of vehicle arrival rate.
As the entry rate increases, all strategies exhibit longer total driving times due to congestion buildup.
The {Customized Solution} consistently yields the shortest times, while the {Data-Driven} approach closely follows.
The {Closest} strategy performs well at low arrival rates but deteriorates quickly as inflow intensifies, whereas {Random} remains the least efficient at light traffic because it fails to exploit nearby available spots.

Overall, the results demonstrate that the {Data-Driven} strategy generalizes well across traffic levels and closely approximates the idealized {Customized Solution}, offering a strong balance between efficiency and practical deployability.

\subsection{Simulation Scenario: Pre-occupied Parking Lot} \label{sec:fulldataset} 

     
        

\begin{figure}
     \centering
     \includegraphics[width=.7\linewidth]{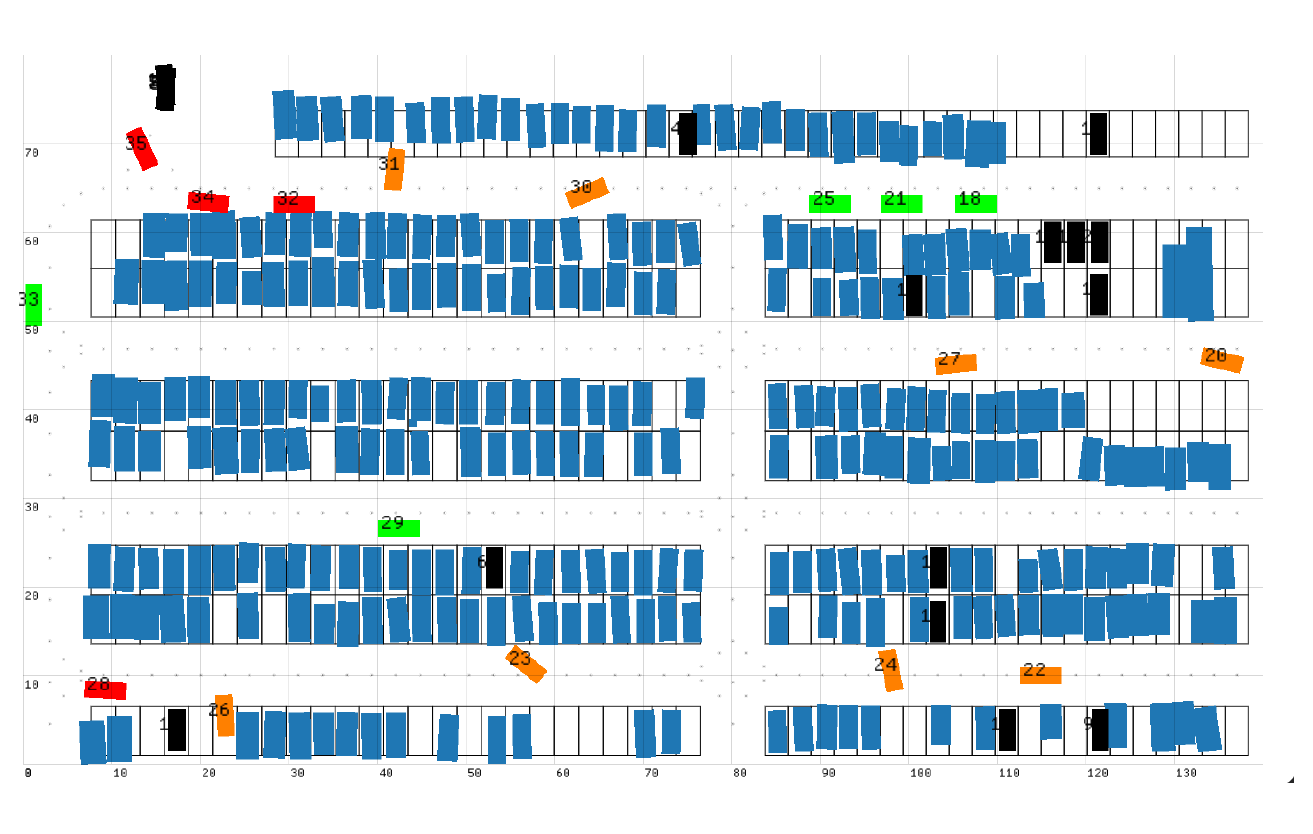}
     \caption{Snapshot of the simulation in a partially occupied parking lot.
    Vehicles marked in blue denote pre-occupied parking spots from the DLP dataset.}
    \label{fig:simulation_validation}
\end{figure}

This scenario provides a realistic comparison between autonomous driving with the proposed parking assignment strategies and the status quo of human driving behavior.
The simulation is initialized with a partially occupied parking lot, where the positions of parked vehicles follow the recorded states in the DLP dataset.
Entering and exiting behaviors, including vehicle arrival times and departure schedules, are also taken directly from the dataset to reflect the actual drivers' parking schedule.

Figure~\ref{fig:simulation_validation} shows a representative snapshot of the simulation.
Unlike the empty-lot scenario (Fig.~\ref{fig:sim_demo_empty_lot}), some parking spaces are already occupied and are shown in blue.
A complete video demonstration of the simulation is available online\footnote{\url{https://youtu.be/qUWVLN-5RSU}}. 

\subsubsection{Methodology}  
We use vehicles $\Agents$ and obstacles $\Obs$ extracted from Scenes 12–30 of the DLP dataset.
Entering vehicles ($\Aenter$) are generated at the lot entrance following their recorded arrival times, while exiting vehicles ($\Aexit$) initiate their unparking maneuvers according to the timestamps observed in the dataset.
For each vehicle $i \in \Agents$, we set its reference cruising speed $v\ofi_{\mathrm{cruise}}$ to the maximum speed observed in the dataset.
Since human drivers exhibit frequent stops and low average speeds, the maximum speed more accurately reflects their intended driving pace.

We evaluate four assignment strategies: {Human Selection, Random, Closest, and Data-driven}.
The Human Selection strategy serves as the baseline, representing actual driver behavior under the limited information of first-person perception.
The Customized Solution strategy is not applicable here, as manual optimization becomes intractable with pre-existing occupancy.
Each scene and strategy combination is simulated 10 times.
 
\begin{figure}
    \centering
    \includegraphics[width = .8\linewidth]{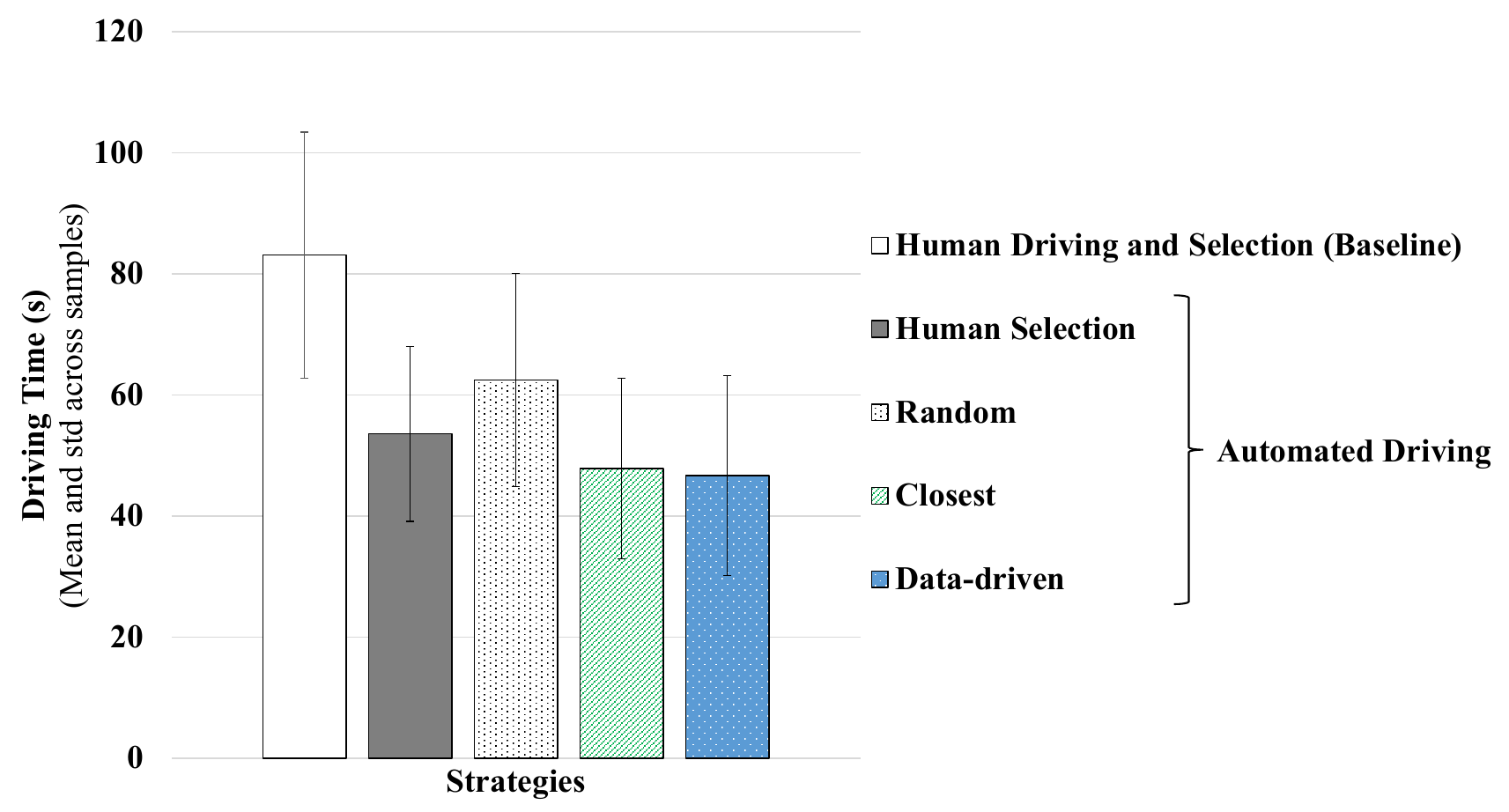}
    \caption{Driving time under different assignment strategies in a partially occupied parking lot based on DLP dataset scenes.
    Automated driving significantly reduces total driving time compared with human driving.
    The Closest and Data-driven strategies outperform Random, with the Data-driven method achieving a 43.8\% reduction from the human baseline.}
    \label{fig:fulldataset}
\end{figure}

\begin{figure}
\centering
\includegraphics[width = .7\linewidth]{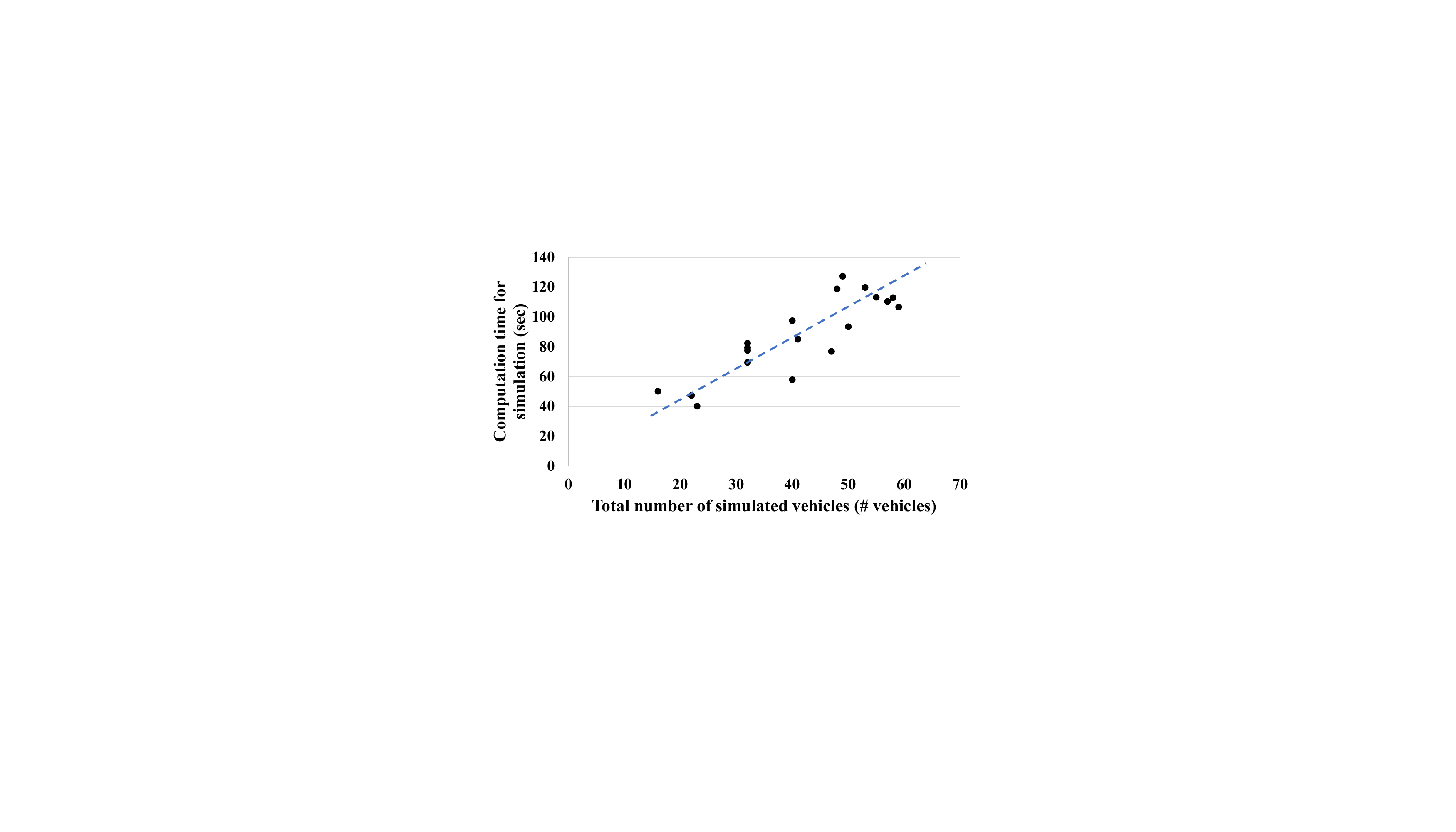}
\caption{Computation time of the simulator versus the total number of simulated vehicles, evaluated using Scenes 12–30 from the DLP dataset.}
\label{fig:computationtime}
\end{figure}

\subsubsection{Results}

Figure~\ref{fig:fulldataset} summarizes the performance comparison, where the ``Human Driving and Selection'' case reflects the human-driving behavior in the real-world DLP dataset, while the other four strategies incorporate autonomous driving control elaborated in this paper.
Overall, automation substantially reduces total driving time due to smoother trajectories, continuous speed control, and more efficient collision avoidance.
In contrast, human drivers in the DLP dataset often exhibit inefficient cruising, frequent stop-and-go movements, and suboptimal parking maneuvers.

Among the automated strategies, {Random} yields the highest driving time due to uncoordinated and spatially inefficient assignments.
Both {Closest} and {Data-driven} perform substantially better, with the latter achieving the lowest average driving time—a 43.8\% reduction relative to the human baseline.
While the {Closest} strategy performs reasonably well under light traffic by minimizing individual travel distance, it lacks awareness of global traffic interactions.
Because all vehicles tend to converge near the entrance, this strategy often creates localized congestion and queuing, particularly in denser conditions (as discussed in Sec.~\ref{sec:empty_lot_result}).
In contrast, the {Data-driven} estimator learns from real traffic patterns to balance spot proximity with spatial dispersion, dynamically assigning vehicles in ways that prevent bottlenecks and reduce overall driving time across the fleet.

In addition, Figure~\ref{fig:computationtime} reports the simulation computation time for Scenes 12–30 using the Human Selection assignment with autonomous vehicle control.
For instance, Scene 12 contains 32 vehicles and requires approximately 79.4 seconds to simulate, while the corresponding real-world recording lasts 431.7 seconds on average (standard deviation: 49.7s).
As expected, computation time increases with the number of vehicles simulated, demonstrating near-linear scalability.

\section{Conclusion and Future Work}

This work presents a novel simulation framework for efficiently modeling and analyzing fleets of connected autonomous vehicles operating within a parking lot.
The framework enables the development and evaluation of multi-level strategies—ranging from individual vehicle control and collision avoidance to high-level fleet coordination and parking spot assignment.
Within this environment, we design and compare multiple assignment strategies aimed at improving parking efficiency through automation and communication.
Although the objective functions of these strategies can be generalized, our experiments focus on minimizing total driving time as a key performance metric, which also correlates with reduced energy consumption and improved infrastructure utilization.
Using the proposed multi-vehicle simulator and the Dragon Lake Parking (DLP) dataset, we demonstrate that automation and communication significantly reduce total driving time.
In particular, our Data-driven assignment strategy achieves up to a 43.8\% reduction in parking time compared to the human baseline.

There remain several promising directions for future research.
First, the findings from these simulation-based experiments should be validated through real-world testing with autonomous vehicles.
A gradual transition—from pure simulation to hardware-in-the-loop and simulation-in-the-loop experiments—would be a practical approach, given the economic, technological, and safety challenges of coordinating multiple autonomous vehicles in a full-scale parking lot.

Second, as autonomous vehicles are gradually introduced into everyday traffic, they will coexist with human-driven and partially automated vehicles.
Understanding how such mixed autonomy systems behave is crucial.
Future work should address the control and communication challenges that arise in interactions among autonomous, human-driven, and unconnected vehicles, as well as between vehicles and pedestrians.

Third, as the automotive sector continues to electrify, parking lots are becoming key hubs for both mobility and energy management.
Vehicle batteries are now digitally connected, enabling the integration of charging operations into parking management systems.
Future research could explore joint optimization of parking and charging infrastructure to enhance energy efficiency, maximize the use of renewable energy, and reduce facility costs, ultimately contributing to a more sustainable urban mobility ecosystem.


\section*{Disclosure statement}

The authors declare that there is no conflict of interest regarding the publication of this article.

\section*{Funding}
This research work presented herein is funded by the Advanced Research Projects Agency-Energy (ARPA-E), U.S. Department of Energy under DE-AR0000791.

\appendix 
\label{sec:appendix}

\section{Reformulation of collision avoidance constraints} ~\label{app:obca}

Here, we present the differentiable reformulation of the collision avoidance contraints~\eqref{eq:CA-constr}
\begin{equation*}
\mathrm{dist}\left(\mathbb{B}(z_{k|t}), \mathbb{B}^{[o]}_{\mathrm{obs}}\right) \geq d_{\mathrm{min}}, \forall o.
\end{equation*}

According to~\cite{obca, zhang_autonomous_2019, firoozi}, \eqref{eq:CA-constr} is equivalent to
\begin{subequations}
\label{eq:obca-obs}
\begin{align}
    \forall o \in \mathcal{O}, \exists \lambda^{[o]} \geq 0, \mu^{[o]} \geq 0, s^{[o]} \ : \|s^{[o]}\| & \leq 1, \\
    - g(z_{k|t})^{\top}\lambda^{[o]} - b^{[o], \top} \mu^{[o]} & \geq d_{\mathrm{min}}, \\
    G(z_{k|t}) \lambda^{[o]} + s^{[o]} & = 0, \\
    A^{[o]} \mu^{[o]} - s^{[o]} & = 0,
\end{align}
\end{subequations}
where $G(z_{k|t}), g(z_{k|t})$ describe the vehicle body polygon $\mathbb{B}(z_{k|t})$ when the vehicle states are $z_{k|t}$, and $A^{[o]}, b^{[o]}$ describe the polygon $\mathbb{O}^{[o]}$ of the static obstacles.

\section{Rule-Based Collision Avoidance Pseudocode} \label{app:collavoid}

Here, we present helper functions for the rule-based collision avoidance algorithm introduced in Section \ref{sec:collision-avoidance}.

Algorithm \ref{alg:crashriskassessment} determines which vehicles that a vehicle $i$ will collide with within a certain number of timesteps. It does this by simulating all vehicles' motion using their planned trajectories (lines 5-9) and comparing their polytopes (lines 10-14). $d_\mr{crash\_check}$ is the radius in which we check for possible collisions, and $t_{\mr{look\_ahead}}$ is the number of timesteps we simulate forward to check for collisions.

\begin{algorithm}
    \caption{Collision Risk Assessment}
    \label{alg:crashriskassessment}
    \begin{algorithmic}[1]
        \State \Comment{Determine any vehicle $j$ will collide with $i$ within $t_{\mr{look\_ahead}}$ time steps if all vehicles continue on their current trajectories}
        \State \Comment{Constants: $d_\mr{collision\_check}$,  $t_\mr{look\_ahead}$}
        \State $s_{\mr{crash}}\leftarrow \emptyset$
        \State $s_{\mr{nearby}} \leftarrow$ vehicles within $d_{\mr{collision\_check}}$ of $i$
        \For{$t=1,2,\ldots,t_{\mr{look\_ahead}}$}
            \State Simulate $i$ moving forward one time step
            \For{$j$ in $s_{\mr{nearby}}$}
                \State Simulate $j$ moving forward one time step
            \EndFor
            \For{$j$ in $s_{\mr{nearby}}$}
                \If{$\mathbb{B}\ofi \bigcap \mathbb{B}^{[j]} \neq \emptyset$} \Comment{The vehicle bodies of $j$ and $i$ intersect}
                    \State $s_{\mr{crash}}\leftarrow s_{\mr{crash}}\cup j$ 
                \EndIf
            \EndFor
        \EndFor
        \State \Return $s_{\mr{crash}}$
    \end{algorithmic} 
\end{algorithm}

\begin{figure}
\centering
\includegraphics[width = .7\linewidth]{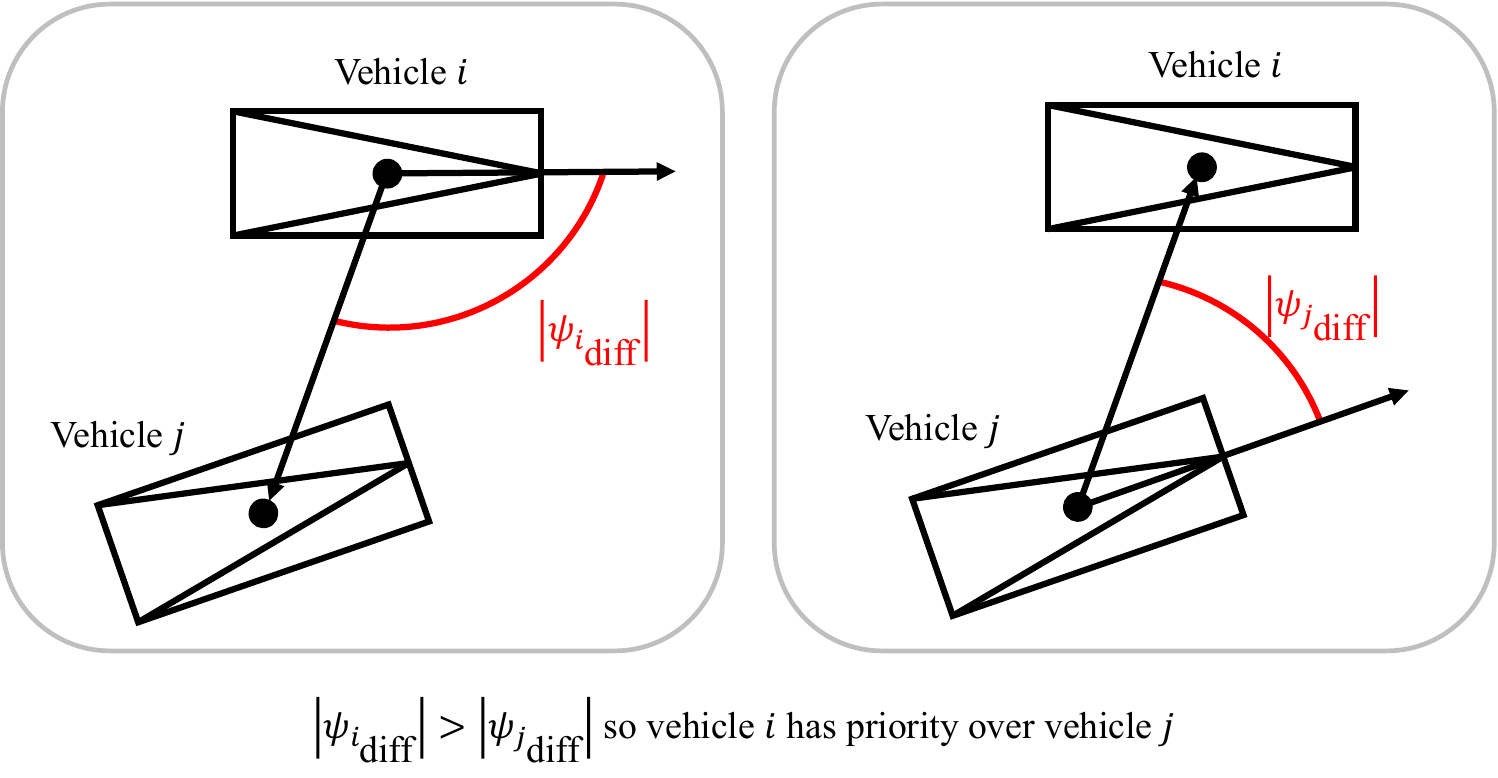}
\caption{Visualization of Algorithm \ref{alg:yieldassessment}, where vehicle $i$ has priority over vehicle $j$}
\label{fig:shouldgobefore}
\end{figure}

Algorithm \ref{alg:yieldassessment} determines priority when two vehicles $i$ and $j$ may collide. It does this using the angle formed by the vector between the two vehicle centers, since this can be used as a measure of which vehicle has gone further past the other and should therefore have priority. A visualization of this algorithm is presented in Figure \ref{fig:shouldgobefore}.

\begin{algorithm} 
    \caption{Priority Assessment}
    \label{alg:yieldassessment}
    \begin{algorithmic}[1]
        \State \Comment{Determine which vehicle will have priority when there is collision risk between vehicle $i$ and $j$}
        \State $\psi_{ij}\leftarrow$ heading angle formed by vector from $i$ to $j$
        \State $\psi_{ji}\leftarrow$ heading angle formed by vector from $j$ to $i$
        \State $\psi_{i_{\mr{diff}}}=\psi_{ij}-\psi_i$, adjusted by factors of $2\pi$ so $|\psi_{i_{\mr{diff}}}|<\pi$
        \State $\psi_{j_{\mr{diff}}}=\psi_{ji}-\psi_j$, adjusted by factors of $2\pi$ so $|\psi_{j_{\mr{diff}}}|<\pi$
        \If{$|\psi_{i_{\mr{diff}}}|>|\psi_{j_{\mr{diff}}}|$}
            \State \Return $i$ has priority
        \Else
            \State \Return $j$ has priority
        \EndIf
    \end{algorithmic} 
\end{algorithm}

\begin{figure}
\centering
\includegraphics[width = .7\linewidth]{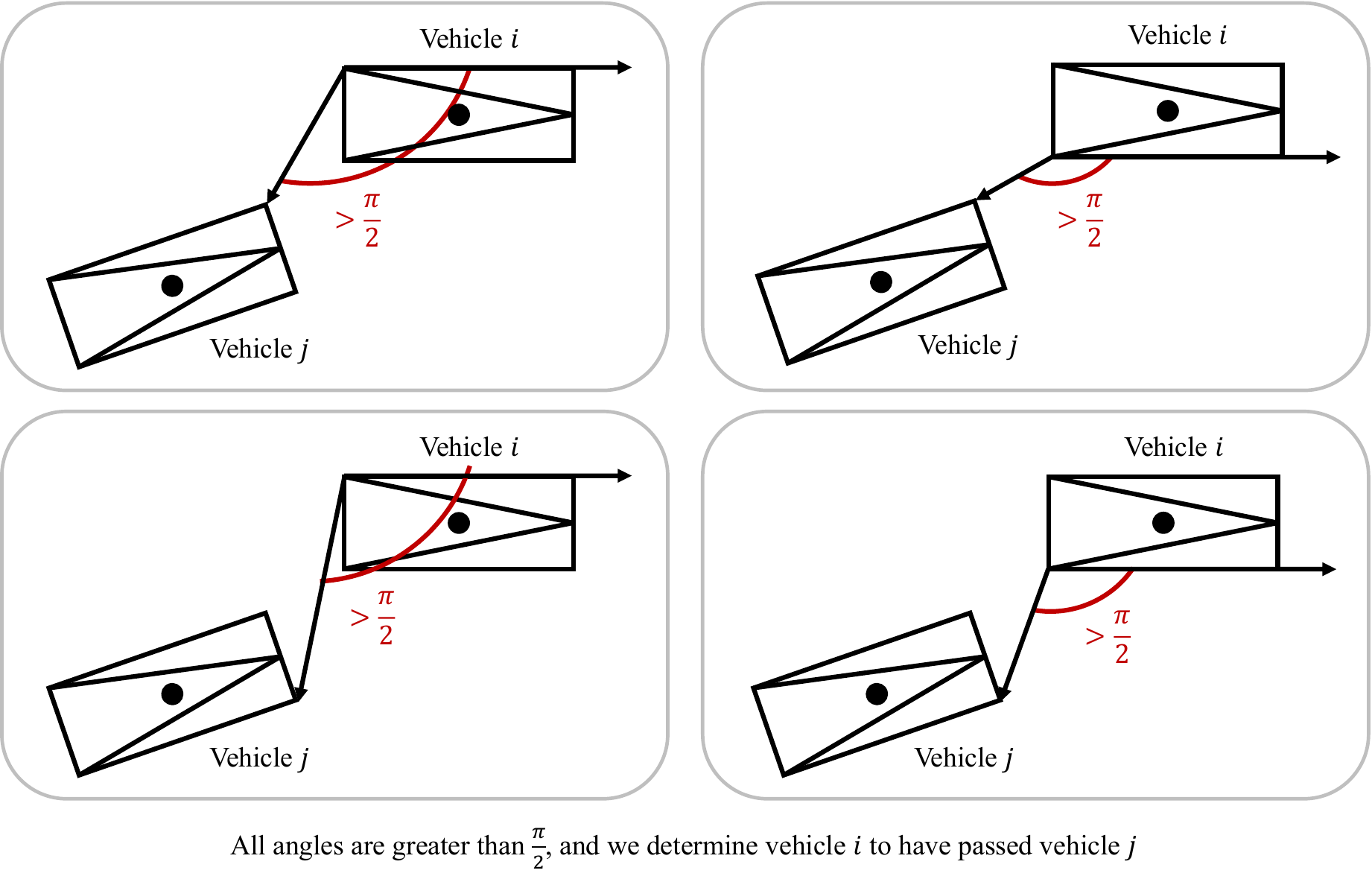}
\caption{Visualization of Algorithm \ref{alg:passage_assessment}, where vehicle $i$ has passed vehicle $j$}
\label{fig:haspassed}
\end{figure}

Algorithm \ref{alg:passage_assessment} determines if a vehicle $i$ has ``driven past" another vehicle $j$, which we define as having the vectors formed by each of its rear corners to each of the front corners of $j$ be more than 90 degrees from $i$'s heading (lines 2-6). There is also an optional buffer distance $d_{\mathrm{buffer}}$, where the center of vehicle $i$ has to be away from $j$ by a certain distance (line 7). A visualization of this algorithm is presented in Figure \ref{fig:haspassed}.

\begin{algorithm} 
    \caption{Passage Assessment}
    \label{alg:passage_assessment}
    \begin{algorithmic}[1]
        \State \Comment{Determine if vehicle $i$ has driven past vehicle $j$ with optional criteria with $d_\mr{buffer}$}
        \For {$c_1\in$ rear corners of $i$}
            \For {$c_2\in$ front corners of $j$}
                \State $\psi_{ij}\leftarrow$ angle formed by vector from $c_1$ to $c_2$
                \If {$|\psi\ofi-\psi_{ij}|<\frac{\pi}{2}$} 
                    \State \Return False
                \EndIf
            \EndFor
        \EndFor
        \If{$d_\mr{buffer}$ is provided and $\sqrt{(x_j-x_i)^2 + (y_j-y_i)^2}<d_\mr{buffer}$} \Comment{$i$ is not $d_\mr{buffer}$ away from $j$}
            \State \Return False
        \EndIf
        \State \Return True
    \end{algorithmic} 
\end{algorithm}

\end{document}